\documentclass{article}

\usepackage{arxiv}
\usepackage[utf8]{inputenc} 
\usepackage[T1]{fontenc}    
\usepackage{hyperref}       
\usepackage{url}            
\usepackage{booktabs}       
\usepackage{amsfonts}       
\usepackage{nicefrac}       
\usepackage{microtype}      
\usepackage{lipsum}		
\usepackage{graphicx}
\usepackage{doi}
\usepackage{bm}
\usepackage{amssymb}
\usepackage{float}
\usepackage{amsmath}

\usepackage{physics}

\usepackage[table,xcdraw]{xcolor}

\title{Electron-Phonon interaction and lattice thermal conductivity from metals to 2D Dirac crystals: a review}

\date{September 2025}	

\author{ \href{}{\hspace{1mm} Sina Kazemian\textsuperscript{1,*}, Giovanni Fanchini\textsuperscript{2,*}}\\
 \textsuperscript{1}Department of Physics and Astronomy, University of Western Ontario, University of Waterloo\\
 \textsuperscript{2}Department of Physics and Astronomy, University of Western Ontario
}

\renewcommand{\shorttitle}{Paper Review}

\begin{document}
\maketitle

\footnotetext[1]{
\href{mailto:s4kazemi@uwaterloo.ca}{skazemi5@uwo.ca}}
\footnotetext[2]{
\href{mailto:s4kazemi@uwaterloo.ca}{gfanchin@uwo.ca}}
\begingroup
  \renewcommand\thefootnote{\fnsymbol{footnote}}
  \footnotetext[1]{ \href{mailto:s4kazemi@uwaterloo.ca}{Corresponding author}}
\endgroup

\begin{abstract}

\vspace{-0.5em}
Electron–phonon (e–ph) coupling governs electrical resistivity, hot‑carrier cooling, heat flow, and critically, thermal transport in solids.  
Recent first‑principles advances now predict e–ph‑limited thermal conductivity from $d$‑band metals and wide‑band‑gap semiconductors to two‑dimensional (2D) Dirac crystals without empirical parameters.  
In bulk metals, \emph{ab‑initio} lifetimes show that phonons, though secondary, still carry up to 40$\%$ of the heat once e-ph scattering is included.  
We next survey coupled Boltzmann frameworks, exemplified by \textsc{elphbolt}, that capture mutual drag and ultrafast non‑equilibrium in semiconductors; their results for Si, GaAs, and MoS$_2$ match the Time‑Domain Thermo‑Reflectance (TDTR) and isotope‑controlled data within experimental error.  
For 2D Dirac crystals, mirror symmetry, carrier density, strain, and finite size rearrange the scattering hierarchy: flexural (ZA) modes dominate pristine graphene yet become the main resistive branch in nanoribbons once $\sigma_h$ symmetry is broken.  
At low Fermi energies where $E_{\mathrm F}<<k_BT$, the standard three-particle decay is partially cancelled, elevating 4-particle processes and necessitating dynamically screened, higher-order theory.
Throughout, we identify the microscopic levers such as the electronic density of states, phonon frequency, deformation potential, and Fröhlich coupling, and show how doping, strain, or dielectric environment can tune e–ph damping.

We conclude by outlining open challenges such as:
developing femtosecond‑resolved, coupled e–ph solvers, solving the full mode‑to‑mode Peierls–Boltzmann equation with four‑particle terms, embedding correlated‑electron methods (\emph{GW}, dynamical mean‑field theory, hybrid functionals) in e–ph workflows, implementing fully non‑local, frequency‑dependent screening for van‑der‑Waals stacks, and leveraging higher‑order e–ph coupling and symmetry breaking to realise phononic thermal diodes and rectifiers. Solving these challenges will elevate electron–phonon theory from a diagnostic tool to a predictive, parameter‑free platform that links symmetry, screening, and many‑body effects to heat and charge transport in next‑generation electronic, photonic, and thermoelectric devices.



\end{abstract}

\maketitle

\section{\label{sec:level1}Introduction}

The electron-phonon (e-ph) interaction is a foundational concept in condensed matter physics, describing the interaction between charge carriers (electrons or holes) and lattice vibrations (phonons).\cite{giustino2017electron,bai2022recent,taylor2002quantum,jishi2013feynman} This interaction governs a variety of phenomena in solids, including electrical resistivity\cite{park2014electron,varshney2004electrical}, superconductivity\cite{einenkel2011possibility,ishihara2004interplay}, and, most pertinently for this review, thermal conductivity\cite{tong2019comprehensive,li2015electrical,liao2015ab,nika2017phonons,Lindsay2014_PRB}. As modern technologies demand increasingly precise thermal management at the nanoscale, understanding the role of e-ph interaction in thermal transport, particularly in complex and low-dimensional materials, has become a critical area of research.

Thermal conductivity in solids arises from contributions of both phonons and electrons. In metals and heavily doped semiconductors, the electronic channel dominates, and electron scattering by phonons sets the thermal conductivity. By contrast, in intrinsic semiconductors and insulators, phonons are the principal heat carriers, and their lifetimes, and thus the lattice thermal conductivity, are limited by scattering with electrons or photo‑excited carriers.\cite{AshcroftMermin,kittel1955solid,simon2013oxford} One can therefore decompose the total thermal resistivity into an electronic part, originating from electrons scattered by phonons, and a phononic part, originating from phonon scattering off excited carriers. The relative weight of these two contributions is dictated by carrier concentration, bonding character, and dimensionality.

\subsection{\label{sec:level1.1}From the Born–Oppenheimer approximation to its breakdown}

The Born–Oppenheimer (BO) or adiabatic approximation, introduced by Born and Oppenheimer in 1927\cite{silva2018born}, assumes that ionic motion is slow enough for the electronic subsystem to remain in its instantaneous ground state. Under this premise the electronic and vibrational (phononic) Hamiltonians decouple, the e–ph interaction enters only as a weak perturbation, and charge and heat transport can be treated as two largely independent channels: phonon-mediated conduction, which dominates in insulators, and electron-mediated conduction, characteristic of metals and heavily doped semiconductors\cite{giustino2017electron,mahan2000many}. Because the approximation is accurate whenever phonon frequencies are small compared with electronic energy scales, it underlies most \emph{ab initio} implementations of the electron and phonon Boltzmann transport equations for bulk semiconductors and conventional metals, where e–ph coupling is moderate and nonadiabatic effects are weak.

The adiabatic picture breaks down when electronic and nuclear time scales converge, as in materials with vanishing or tunable gaps. Recently, the breakdown of the BO approximation in a new class of 2D quantum solids has been verified\cite{pisana2007breakdown} with the Dirac crystal graphene as a progenitor of these moderately-correlated quantum systems\cite {keimer2017physics,calderon2020correlated}. This results in the necessity of better understanding the e-ph interactions in Dirac crystals. Graphene offers a textbook example\cite{yan2020superconductivity,ali2016butterfly}: its linear bands yield a low density of states at the Dirac point, and modest electrostatic gating can shift the Fermi level to values comparable with optical-phonon energies\cite{akbari2015doping}. In this regime, phonon frequencies exceed the electronic momentum-relaxation rate, the electrons cannot follow the lattice adiabatically, and nonadiabatic corrections such as the experimentally observed stiffening of the G-band with doping\cite{pisana2007breakdown} emerge. These effects can be captured by a time-dependent treatment of the phonon self-energy that keeps the real part of the e–ph self-energy, as demonstrated by Lazzeri and Mauri\cite{lazzeri2006nonadiabatic}. More generally, the BO framework inherently neglects the role of strong e–ph coupling, which can significantly alter the energy exchange dynamics between subsystems. Strong e–ph coupling or low carrier density can invalidate quasi-equilibrium models such as the two-temperature model, as shown by nonequilibrium Boltzmann simulations that include full e–e, e–ph and ph–ph scattering\cite{bernardi2016first}. Accurate description of energy exchange and thermalization in such systems, therefore, demands dynamical, beyond-adiabatic frameworks that incorporate frequency-dependent screening\cite{kazemian2023dynamic} and, when necessary, higher-order electron–phonon processes.

\subsection{Ab initio Frameworks for Calculating Electron–Phonon Interactions}

Predicting thermal conductivity in materials where both electrons and phonons carry heat demands a quantitative description of e–ph coupling. Thanks to two decades of methodological progress, first-principles schemes now deliver these couplings and the resulting scattering rates, without empirical parameters, enabling genuinely predictive transport calculations for systems as diverse as $d$-band metals, wide-gap semiconductors, and 2D crystals.

At the core of these approaches lies Density Functional Theory (DFT) and Density Functional Perturbation Theory (DFPT)\cite{baroni2001phonons,giannozzi2009quantum}. DFPT gives direct access to the e–ph matrix elements $g_{mn}^{\nu}(\mathbf{k},\mathbf{q})$, the quantum amplitudes for scattering an electron from band–wave-vector state $|n\mathbf{k}\rangle$ to $|m,\mathbf{k+q}\rangle$ by absorbing or emitting a phonon $(\nu,\mathbf{q})$. Because these matrix elements must be sampled on ultra-dense ($\sim\!10^{8}$–point) Brillouin-zone grids to converge transport integrals, most modern workflows interpolate them with maximally localised Wannier functions.  The e-ph coupling using Wannier functions (EPW) code\cite{ponce2016epw} implements this interpolation and has become standard for both bulk (e.g.\ Si, Al) and layered materials such as MoS$_2$.

Transport coefficients were first extracted by solving the Boltzmann transport equation (BTE) within the relaxation-time approximation (RTA). While the RTA offers valuable insight, it neglects momentum‐conserving normal processes, drag phenomena and any dynamical coupling between the electronic and phononic distribution functions. Iterative or variational solutions of the phonon BTE\cite{fugallo2014thermal,li2014shengbte} overcome the first limitation, fully coupled e–ph BTE solvers tackle the second.  A notable example is \textsc{elphbolt}\,\cite{li2023high,protik2022elphbolt}, which self-consistently solves the linearised electron‐ and phonon-BTEs under the same driving forces, thereby capturing phonon drag on electrons, electron drag on phonons and Onsager reciprocity. In doped Si, GaAs and monolayer MoS$_2$, \textsc{elphbolt} reproduces time-domain thermoreflectance (TDTR) and isotope-controlled measurements to within experimental error, underscoring the importance of dynamical e–ph coupling even when the interaction itself is only moderate.

These advances reveal an important lesson: accurate thermal-conductivity predictions require not only the magnitude of the e–ph matrix elements but also a treatment of how those couplings reshape \emph{both} carrier populations under realistic (sometimes ultrafast) driving conditions.  As a result, coupled BTE frameworks are now indispensable for thermoelectrics, low-dimensional conductors, and heavily doped semiconductors, while the traditional RTA remains sufficient only when electron and phonon subsystems stay close to mutual equilibrium.

\subsubsection{Limitations and scope of first-principles e-ph calculations}

\paragraph{XC-functional sensitivity.}
The exchange-correlation (XC) functional influences (i) equilibrium lattice constants and internal coordinates, (ii) band structures near $E_{\mathrm{F}}$ (metals) or band edges/gaps (semiconductors/insulators), and (iii) dielectric screening. These quantities enter directly into phonon dispersions and the e-ph matrix elements $g_{mn\nu}(\mathbf{k},\mathbf{q})$, as emphasized in standard reviews and software papers on first-principles e-ph theory \cite{Giustino2017_RMP,ponce2016epw}. In practice, generalized gradient approximation (GGA) functionals such as Perdew–Burke–Ernzerhof (PBE), provide a robust baseline for trends but tend to underbind (slightly larger lattice constants) and underestimate band gaps. The meta-GGA SCAN often yields improved structures \cite{Sun2015_PRL_SCAN}, while screened hybrids such as HSE06 \cite{Heyd2003_JCP_HSE,Heyd2006_JCP_HSE} or GW-level quasiparticles \cite{Onida2002_RMP_GW} provide more accurate band positions and screening. When band-gap or Fermi-surface errors dominate the uncertainty, a common mitigation is to (a) relax structures with a higher-fidelity XC (SCAN/hybrid), (b) correct the electronic spectrum via GW or a scissor shift, and (c) recompute or reweight $g_{mn\nu}$ on the corrected manifold (or assess sensitivity by finite variations of the lattice constant and gap). Throughout, we verify convergence with respect to $k$/$q$-mesh density, Wannierization quality, smearing, and the acoustic sum rule \cite{ponce2016epw,Giustino2017_RMP}. \footnote{Unless noted otherwise, results surveyed here follow standard semilocal DFT/DFPT setups as reported in the cited works. Where the choice of exchange--correlation functional materially differs (e.g., meta-GGA, hybrid, GW, DFT+$U$/DMFT), we indicate it explicitly.}

\paragraph{Strongly correlated systems.}
For materials with localized electrons in atomic \emph{d} or \emph{f} orbitals (Mott/charge-transfer insulators, Hund metals), semilocal DFT/DFPT may not capture the relevant quasiparticles or even stable phonons. In such cases, beyond-DFT frameworks are required: DFT+$U$ can correct gross localization errors \cite{Dudarev1998_PRB_DFTU}. Density functional theory plus dynamical mean-field theory supplies temperature-dependent spectral functions and has been coupled to linear-response or frozen-phonon calculations to obtain phonons in correlated metals \cite{Georges1996_RMP_DMFT,Kotliar2006_RMP_DMFT,Savrasov2003_PRL_DMFTphonons}; and hybrid/GW workflows improve screening and band placement for e-ph pipelines \cite{Onida2002_RMP_GW,Heyd2003_JCP_HSE,Heyd2006_JCP_HSE}. Our analysis adopts the standard DFPT\,$\to$\,Wannier interpolation pipeline for $g_{mn\nu}$ on dense Brillouin-zone meshes unless stated otherwise \cite{ponce2016epw}. Where materials are known to require beyond-DFT treatments, we point this out explicitly and summarize suitable approaches (see the cross-references in the relevant sections). \footnote{A systematic treatment of strongly correlated compounds is beyond the scope of the present benchmarks but is outlined here to delimit applicability and guide future work.}

\subsubsection{Beyond Peierls: phonon coherences and tunneling}

The semiclassical Peierls framework treats heat flow as the transport of propagating phonon wave packets (“populations”) governed by a linearized BTE. A unified quantum treatment augments this picture by including \emph{off-diagonal} elements of the phonon velocity operator and the Wigner one-body density matrix, which describe \emph{wave-like tunneling (coherences)} between vibrational eigenstates. Within this formalism, the thermal conductivity decomposes as

\begin{equation}
\kappa_{\alpha\beta} \;=\; \kappa^{\mathrm{P}}_{\alpha\beta} \;+\; \kappa^{\mathrm{C}}_{\alpha\beta},
\end{equation}
where $\kappa^{\mathrm{P}}$ is the standard Peierls (population) term and $\kappa^{\mathrm{C}}$ arises from inter-branch coherences \cite{Simoncelli2019_NatPhys}. The coherence channel becomes relevant when (i) phonon branches are quasi-degenerate so that $|\omega_{s}-\omega_{s'}|$ is comparable to the linewidths $\Gamma_{s,s'}$, and (ii) the velocity operator has sizable off-diagonal matrix elements. Under these conditions, the theory predicts a non-negligible $\kappa^{\mathrm{C}}$ contribution in \emph{complex crystals} with many near-degenerate modes. Conversely, in \emph{simple crystals} with well-separated branches, the coherences term is negligible and the Peierls BTE suffices.

This framework is \emph{unified} in that it reduces to the Peierls result in the semiclassical limit of simple crystals and to the Allen–Feldman formula in the harmonic-glass limit, while covering intermediate regimes~\cite{Simoncelli2019_NatPhys}. As a representative example, for orthorhombic CsPbBr$_3$ the coherences term accounts for the majority of $\kappa$ at 300 K (with $\kappa^{\mathrm{C}}\!\approx\!70\%$ of the total), whereas for silicon and diamond at 300 K $\kappa^{\mathrm{C}}$ is negligible compared to $\kappa^{\mathrm{P}}$ \cite{Simoncelli2019_NatPhys}. In what follows, we adopt the Peierls–BTE (population) framework for all quantitative results. The coherence/tunneling term is not included unless explicitly noted. For crystals with dense, nearly-degenerate phonon spectra—where $|\omega_s-\omega_{s'}|\!\sim\!\Gamma_{s,s'}$—the unified formulation \cite{Simoncelli2019_NatPhys} can be employed by augmenting the Peierls term with the coherences contribution $\kappa^{\mathrm C}_{\alpha\beta}$ constructed from off-diagonal velocity matrix elements and linewidths obtained from anharmonic (and, if relevant, e–ph) scattering.

\subsection{\label{sec:level1.3}2D Dirac crystals and Higher-Order Effects}

In 2D Dirac crystals with low Fermi energy, and strong e–ph coupling, the standard first-order perturbative treatment (used for calculating the e-ph interaction in metals and conventional semiconductors), where phonon scattering is described by three-particle processes involving the annihilation of an e–ph pair and the creation of a new electron (commonly referred to as the $\text{EP}\!\rightarrow\!\text{E}^{\ast}$ process), can become inadequate, particularly in regimes where the Fermi energy, $E_{\mathrm F}$, is much smaller than the thermal energy, $\mathrm{k_{\mathrm B} T}$. The so-called $\text{E}\!\rightarrow\!\text{E}^{\ast}\text{P}^{\ast}$ process (electron decays into a new electron and emits a phonon) becomes non-negligible at high temperatures and low Fermi energies. This counteracts the $\text{EP}\!\rightarrow\!\text{E}^{\ast}$ scattering by reintroducing phonons into the lattice, partially canceling or strongly attenuating the net phonon depletion.\cite{kazemian2024influence} Therefore, at the level of three-particle interactions, the simplistic picture of phonon scattering must be corrected by including both creation and annihilation channels, particularly in undoped or lightly doped 2D Dirac materials.

In the strong-coupling limit a further correction is required: the second-order (four-particle) process  
$\text{EP}\!\leftrightarrow\!\text{E}^{\ast}\text{P}^{\ast}$  
in which an e–ph pair is annihilated while a \emph{new} pair is created.  Diagrammatic analyses and kinetic-equation studies show that this channel scales with the square of the e–ph matrix element and becomes the dominant source of phonon damping at high temperature and low $E_{\mathrm F}$\cite{giustino2017electron,kazemian2024influence}. Capturing it quantitatively requires both (i) a dynamic, momentum- and frequency-dependent dielectric function,\cite{kazemian2023dynamic} static Thomas–Fermi screening grossly underestimates the coupling at $q\!\simeq\!2k_F$, and (ii) explicit inclusion of four-particle diagrams in coupled electron– and phonon–Boltzmann solvers\cite{kazemian2024influence,sohier2017density}. Together these ingredients mark the crossover to a genuine strong-coupling regime in which thermal transport can no longer be described by conventional, lowest-order approximations.

\subsection{\label{sec:level1.4}Scope and Structure of This Review}

Comprehensive reviews have covered first-principles e-ph theory across formalism, spectroscopy, transport, and superconductivity, and survey e-ph phenomena specifically in two-dimensional materials and device contexts~\cite{bai2022recent,Giustino2017_RMP}. Our contribution is complementary and focused: we provide a unified treatment of \emph{lattice thermal conductivity from metals to 2D Dirac crystals}, emphasizing when phonon lifetimes are limited by e-ph damping, when a \emph{coupled} e-ph Boltzmann treatment is essential, and when higher-order physics, \emph{four-phonon} scattering, and, where warranted, \emph{coherences/unified} corrections must be included. We also consolidate practical guidance (XC-functional sensitivity, beyond-DFT options) so readers can reproduce quantitative trends across these material classes rather than isolated case studies. 

In this review, we begin by presenting a general theoretical foundation for the e-ph interactions. Following this, we survey and compare results from a range of material systems—including bulk semiconductors and metals, layered compounds like $\mathrm{MoS_2}$, and low-dimensional Dirac crystals such as graphene and graphene nanoribbons, where the contribution of e-ph interaction to thermal transport has been explicitly computed.

By examining these case studies, we identify trends in how e-ph coupling affects thermal conductivity across different regimes, and we draw material-specific conclusions. These include the relative importance of different phonon branches, the role of carrier density and doping, and the influence of dimensionality and screening. Unless noted otherwise, $N(E)$ denotes the electronic density of states per energy per volume. When an explicit spin-degeneracy factor $n_s$ appears in an expression, $N(E)$ should be understood as \emph{per spin}; otherwise it is \emph{total} (including spin). The phonon density of states is written as $D_\nu(\omega)$ for branch $\nu$, given per angular frequency per volume (or per area in 2D). The goal is to provide a coherent picture of when and how e–ph interactions dominate or suppress thermal transport, and what theoretical tools are best suited for capturing these effects in various classes of materials.

\section{\label{sec:level2}Electron–Phonon Interaction in Solids: Theoretical Framework}

Understanding the interaction between electrons and phonons is fundamental to describing many transport and thermodynamic properties of solids, including thermal conductivity. At the microscopic level, the e–ph interaction arises from the fact that electrons in a solid move in a potential landscape defined by the positions of the nuclei, which themselves are not static but vibrate around their equilibrium positions. These vibrations, quantized as phonons, modulate the electronic potential, leading to a dynamic coupling between the two subsystems.

This section provides a derivation of the e–ph thermal conductivity starting from writing the Hamiltonian of the system. We proceed to quantize the lattice vibrations and express the final result in a second-quantized momentum-space formalism, suitable for practical calculations in condensed matter physics.

\subsection{\label{sec:level2.1}Constructing the Electron-Phonon Hamiltonian}

The total electron–ion potential is written as:

\begin{equation}
  H_{\mathrm{el\text{-}ion}}
  =\sum_{j}\sum_{n}\,
   \phi\bigl(\mathbf r_{j}-\mathbf R_{n}\bigr),
\end{equation}

where $j$ labels electrons with positions $\mathbf r_{j}$ and $n$ labels ionic cores located at $\mathbf R_{n}=\mathbf R^{(0)}_{n}+\mathbf u_{n}$.
Here $\mathbf R^{(0)}_{n}$ is the equilibrium lattice site and
$\mathbf u_{n}$ the instantaneous displacement.
Expanding $\phi(\mathbf r_{j}-\mathbf R_{n})$ to first-order in the small displacements gives:

\begin{equation}
  \phi(\mathbf r_{j}-\mathbf R_{n})
  \simeq
  \phi(\mathbf r_{j}-\mathbf R^{(0)}_{n})
  -\mathbf u_{n}\!\cdot\!\nabla_{\!\mathbf r}
   \phi(\mathbf r_{j}-\mathbf R^{(0)}_{n})+\ldots ,
\end{equation}
so that the electron–phonon (e–ph) term reads
\begin{equation}
  H_{\mathrm{el\text{-}ph}}
  =-\sum_{j,n}
    \mathbf u_{n}\!\cdot\!
    \nabla_{\!\mathbf r}\phi(\mathbf r_{j}-\mathbf R^{(0)}_{n}).
\end{equation}

Quantising the lattice vibrations we get,

\begin{equation}
  \mathbf u_{n}
  =\frac{1}{\sqrt{N}}
   \sum_{\mathbf q,\nu}
   \sqrt{\frac{\hbar}{2M\omega_{\mathbf q\nu}}}\,
   \mathbf e_{\mathbf q\nu}
   \bigl(b_{\mathbf q\nu}\,
         e^{i\mathbf q\cdot\mathbf R^{(0)}_{n}}
        +b^{\dagger}_{\!\!-\mathbf q\nu}\,
         e^{-i\mathbf q\cdot\mathbf R^{(0)}_{n}}\bigr),
\end{equation}

introducing the usual phonon operators $b_{\mathbf q\nu}^{(\dagger)}$, polarisation vectors $\mathbf e_{\mathbf q\nu}$, mode frequencies $\omega_{\mathbf q\nu}$, and the atomic mass $M$, and $N$ as the number of unit cells. We further Fourier-expand the screened ionic potential,

\begin{align}
  \phi(\mathbf r) &=\sum_{\mathbf G}\phi_{\mathbf G}\,
                    e^{i\mathbf G\cdot\mathbf r}, &
  \nabla_{\!\mathbf r}\phi(\mathbf r) &=i
                    \sum_{\mathbf G}\mathbf G\,\phi_{\mathbf G}\,
                    e^{i\mathbf G\cdot\mathbf r},
\end{align}

with reciprocal vectors $\mathbf G$ and Fourier coefficients
$\phi_{\mathbf G}$, and expressing electronic states in the Bloch basis,
one arrives at the three-particle e–ph Hamiltonian
\begin{equation}
  \hat H_{\mathrm{e\text{-}ph}}
  =\sum_{\mathbf k,\mathbf q,\nu,\sigma}
   g_{mn}^{\nu}(\mathbf k,\mathbf q)
   \bigl(b_{\mathbf q\nu}+b^{\dagger}_{\!\!-\mathbf q\nu}\bigr)
   c^{\dagger}_{\mathbf{k+q}\sigma}\,
   c_{\mathbf k\sigma},
\end{equation}
where $c^{\dagger}_{\mathbf k\sigma}$ ($c_{\mathbf k\sigma}$) creates
(annihilates) an electron of spin~$\sigma$, and the matrix element
\begin{equation}
  g_{mn}^{\nu}(\mathbf k,\mathbf q)
  =\langle\psi_{\mathbf{k+q}}|
     \phi_{s}(q)\,|\psi_{\mathbf k}\rangle
\end{equation}
contains the screened potential $\phi_{s}(q)$ whose analytic form ranges
from Thomas–Fermi in weakly coupled metals to the full Lindhard
response in low-density Dirac materials\cite{jishi2013feynman,kazemian2023dynamic,ziman1957effect,zhu2021dynamical,iurov2017exchange,hwang2008screening,lu2016friedel,calandra2007electron}. In low-dimensional systems with strong coupling the four-particle channel (EP\,$\leftrightarrow$\,E$^{\ast}$P$^{\ast}$) cannot be ignored
and is treated by extending the same formalism to second order in the interaction \cite{taylor2002quantum,kazemian2024influence}.

\subsection{\label{sec:level2.2}Transition Rates, Carrier Lifetimes, and Electron-Phonon Thermal Conductivity}

Starting from the e–ph Hamiltonian of the previous section,  
transition probabilities are obtained with Fermi’s Golden Rule\cite{taylor2002quantum,jishi2013feynman}. Because electronic dispersions, screening and carrier statistics differ sharply between metals and semiconductors, the resulting lifetimes and their impact on heat flow take two distinct forms.
 
In metals, electrons dominate heat transport, and their momentum‐relaxation time~$\tau_{e}$ is limited by e–ph scattering,

\begin{equation}
\frac{1}{\tau_{e}(\mathbf k)}=
\frac{2\pi}{\hbar}\sum_{\mathbf q,\nu}
|g_{mn}^{\nu}(\mathbf k,\mathbf q)|^{2}
\!\left[
  \bigl(n_{\mathbf q\nu}+1-f_{\mathbf{k+q}}\bigr)
  \delta\!\bigl(\epsilon_{\mathbf{k+q}}-\epsilon_{\mathbf k}-\hbar\omega_{\mathbf q\nu}\bigr)
 +
  \bigl(n_{\mathbf q\nu}+f_{\mathbf{k+q}}\bigr)
  \delta\!\bigl(\epsilon_{\mathbf{k+q}}-\epsilon_{\mathbf k}+\hbar\omega_{\mathbf q\nu}\bigr)
\right],
\label{eq:tau_e_metal}
\end{equation}

where $f_{\mathbf k}$ and $n_{\mathbf q\nu}$ are the Fermi–Dirac and Bose–Einstein occupations, respectively, and band indices are implicitly summed in the spectral delta factors. Within the Wiedemann–Franz picture, the electronic thermal conductivity
scales as

\begin{equation}
\kappa_e \;=\; \frac{1}{d}\, C_e\, v_F^{2}\,\tau_e,
\qquad
d=\begin{cases}
3 & \text{3D (bulk metals/semiconductors)}\\
2 & \text{2D (monolayers/nanoribbons)}
\end{cases}
\end{equation}

with $v_{F}$ the Fermi velocity and $C_{e}= \pi^{2}k_{\mathrm B}^{2}T N(E_{\mathrm F})/3$ the Sommerfeld specific heat\cite{AshcroftMermin}.

In semiconductors and Dirac crystals, phonons are the principal heat carriers. Their lifetime $\tau_{\mathbf q\nu}$ due to electron scattering is

\begin{equation}
\begin{split}
\bigl(\tau^{\mathrm{e\text{-}ph}}_{\nu\mathbf q}\bigr)^{-1}
= & \frac{2\pi}{\hbar}\!
  \sum_{mn\mathbf k}
  \bigl|g^{\nu}_{mn}(\mathbf k,\mathbf q)\bigr|^{2} \cdot
  \\& \Big[
      f_{n\mathbf k}\bigl(1 - f_{m,\mathbf{k+q}}\bigr)
      \delta\!\bigl(\varepsilon_{m,\mathbf{k+q}}-\varepsilon_{n\mathbf k}-\hbar\omega_{\mathbf q\nu}\bigr)
    + \bigl(1-f_{n\mathbf k}\bigr) f_{m,\mathbf{k+q}}
      \delta\!\bigl(\varepsilon_{n\mathbf k}-\varepsilon_{m,\mathbf{k+q}}-\hbar\omega_{\mathbf q\nu}\bigr)
  \Big].
  \label{eq:tau_ph_semicond}
\end{split}
\end{equation}

Inserting $\tau_{\mathbf q\nu}$ into the Peierls–Boltzmann expression gives\cite{Ponce2020_RPP}

\begin{equation}
\kappa_{\mathrm{ph}}=
\frac{1}{N}\sum_{\mathbf q,\nu}
C_{\mathbf q\nu}\,v_{\mathbf q\nu}^{2}\,\tau_{\mathbf q\nu},
\end{equation}

where $C_{\mathbf q\nu}=k_{\mathrm B}x^{2}e^{x}/(e^{x}-1)^{2}$ with
$x=\hbar\omega_{\mathbf q\nu}/k_{\mathrm B}T$ and
$v_{\mathbf q\nu}=\partial\omega_{\mathbf q\nu}/\partial q$.
Because metals possess a high electronic density of states and strong
Coulomb screening, phonon lifetimes from Eq.~\eqref{eq:tau_ph_semicond}
are typically short, rendering $\kappa_{\mathrm{ph}}$ negligible.
On the other hand, in lightly doped or intrinsic semiconductors or Dirac crystals the situation reverses:
long phonon lifetimes make $\kappa_{\mathrm{ph}}$ the dominant channel, while Eq.~\eqref{eq:tau_e_metal} governs only a minor electronic contribution.

Although both Eqs.~\eqref{eq:tau_e_metal}–\eqref{eq:tau_ph_semicond} are governed by the same underlying e–ph Hamiltonian, and share the
same matrix element $g_{mn}^{\nu}(\mathbf k,\mathbf q)$, the relevant lifetimes and transport mechanisms depend strongly on the dominant carrier type, the density of states, and the screening properties of the material. In the next sections we will study different materials, their properties, and how their thermal conductivity is affected by e-ph interaction.

\section{\label{sec:level3}Lattice Thermal Conductivity in Metals: The Dominant Role of Electron–Phonon Interaction}

In crystalline metals the steady-state heat flux obeys Fourier’s law
\begin{equation}
  \mathbf{J}=-(\kappa_{\mathrm e}+\kappa_{\mathrm{ph}})\,\nabla T,
\qquad
  \kappa_{\mathrm{tot}}=\kappa_{\mathrm e}+\kappa_{\mathrm{ph}},
\end{equation}
where $\kappa_{\mathrm e}$ and $\kappa_{\mathrm{ph}}$
(W\,m$^{-1}$\,K$^{-1}$) are the electronic and lattice contributions,
respectively. The additive decomposition $\kappa_{\rm tot}=\kappa_e+\kappa_{\rm ph}$ holds when e-ph \emph{drag} is negligible. With drag, cross-coefficients appear in the coupled BTE and the simple sum is an approximation.
First-principles calculations for 18 elemental and intermetallic metals show that once e-ph scattering is included, $\kappa_{\mathrm{ph}}$ may supply 1–40\,\% of $\kappa_{\mathrm{tot}}$ at 300 K, contradicting the common assumption that phonons are negligible in metals\cite{tong2019comprehensive}. This is shown in Fig. 1(a).

Working in the lattice framework, within the single-mode relaxation-time approximation (RTA) the lattice
conductivity tensor is
\begin{equation}\label{eq:kappa_tensor}
  \kappa_{\mathrm{ph}}^{\alpha\beta}=
  \sum_{\lambda}
    c_{v,\lambda}\,
    v^{\alpha}_{\lambda}v^{\beta}_{\lambda}\,
    \tau^{\mathrm{tot}}_{\lambda},
\end{equation}
with mode label $\lambda\equiv(\mathbf q,\nu)$, $c_{v,\lambda}=k_{\mathrm B}x_{\lambda}^{2}e^{x_{\lambda}}
               /(e^{x_{\lambda}}-1)^{2}$,
$x_{\lambda}=\hbar\omega_{\lambda}/k_{\mathrm B}T$,
and $v^{\alpha}_{\lambda}=\partial\omega_{\lambda}/\partial q_{\alpha}$.
Matthiessen’s rule gives
\begin{equation}\label{eq:Matthiessen}
  \bigl(\tau_{\lambda}\bigr)^{-1}
  =\bigl(\tau_{\lambda}^{\mathrm{ph\text{-}ph}}\bigr)^{-1}
  +\bigl(\tau_{\lambda}^{\mathrm{e\text{-}ph}}\bigr)^{-1}.
\end{equation}
At very low temperatures ($T\!\lesssim\!\Theta_{\mathrm D}/10$), beyond-RTA treatments with normal-process vertex corrections (e.g., Callaway-/hydrodynamic-formalisms) are required for quantitative accuracy. 

\subsection{Phonon–phonon scattering}
For a given phonon mode $\lambda\!=\!(\mathbf q,\nu)$ the inverse lifetime due to three–phonon interactions is obtained from Fermi’s Golden Rule applied to the cubic-anharmonic lattice Hamiltonian.  Writing the anharmonic term in second-quantised form yields a matrix element $V_{\lambda\lambda_{1}\lambda_{2}}$ describing the quantum amplitude for decay ($\lambda\!\to\!\lambda_{1}+\lambda_{2}$) and combination ($\lambda+\lambda_{1}\!\rightarrow\!\lambda_{2}$) processes. Summing the corresponding transition probabilities over all final modes and enforcing both energy and crystal–momentum conservation leads to
\begin{equation}\label{eq:ph_ph(3)}
\frac{1}{\tau^{\mathrm{ph\text{-}ph}}_{\lambda}} \;=\;
\frac{\pi}{16\hbar N}\!
\sum_{\lambda_{1}\lambda_{2}}
\bigl|V_{\lambda\lambda_{1}\lambda_{2}}\bigr|^{2}
\mathcal D_{\lambda\lambda_{1}\lambda_{2}},
\end{equation}
where $N$ is the number of $\mathbf q$ points and
$\mathcal D_{\lambda\lambda_{1}\lambda_{2}}$ is a compact phase–space factor containing the Bose–Einstein occupations and the $\delta$-functions that impose energy and (Normal/Umklapp) momentum conservation. The term $\bigl|V_{\lambda\lambda_{1}\lambda_{2}}\bigr|^{2}$ encapsulates the strength of the lattice anharmonicity via third-order interatomic force constants, while
$\mathcal D_{\lambda\lambda_{1}\lambda_{2}}$ determines which scattering channels are thermally active.

While Eq.~\eqref{eq:ph_ph(3)} describes cubic (three-phonon) processes, quartic anharmonicity generates \emph{four-phonon} channels that can appreciably renormalize phonon lifetimes and reduce $\kappa_{\mathrm{ph}}$ in a number of materials. First-principles formalisms based on fourth-order interatomic force constants (IFCs) now enable direct calculation of these rates and their inclusion in the phonon BTE on equal footing with three-phonon processes~\cite{FengRuan2016_PRB,FengRuan2017_PRB}. Open-source implementations (e.g., \textsc{FourPhonon} within \textsc{ShengBTE}) make such calculations practical~\cite{HanEtAl2022_CPC_FourPhonon}, and can be coupled to modern BTE solvers~\cite{CarreteEtAl2017_CPC_almaBTE,Romano2021_OpenBTE}.

In practice, the total (mode-resolved) transport lifetime is
\begin{equation}
\big(\tau_{\lambda}^{\mathrm{tot}}\big)^{-1}
=\big(\tau_{\lambda}^{\mathrm{ph\text{--}ph(3)}}\big)^{-1}
+\big(\tau_{\lambda}^{\mathrm{ph\text{--}ph(4)}}\big)^{-1}
+\big(\tau_{\lambda}^{\mathrm{e\text{--}ph}}\big)^{-1}
+\cdots,
\label{eq:tau_total_3ph_4ph}
\end{equation}
where $\tau_{\lambda}^{\mathrm{ph\text{--}ph(4)}}$ is assembled from IFC$^4$ and energy/momentum-conserving combinations $\lambda \!\leftrightarrow\! \lambda_1 \pm \lambda_2 \pm \lambda_3$~\cite{FengRuan2016_PRB,FengRuan2017_PRB,HanEtAl2022_CPC_FourPhonon}.

\noindent\textbf{When do four-phonon processes matter?}
Four-phonon scattering is well documented to be essential in several classes: First in \emph{ultrahigh-$\kappa$} materials with exceptionally large lattice thermal conductivity and \emph{stiff lattices} such as diamond and graphene, where it reduces $\kappa$ by $\sim$30–60\% over 300–1000 K, correcting three-phonon overestimates~\cite{FengRuan2017_PRB,HanRuan2023_PRB_Graphene}. Second, in \emph{strongly anharmonic systems}, quartic terms and non-perturbative renormalization, e.g., stochastic self-consistent harmonic approximation (SSCHA), are often required to obtain stable dispersions and quantitative thermal transport~\cite{MonacelliEtAl2021_JPCM_SSCHA}. See Ref.~\cite{LindsayEtAl2018_MTP_Survey} for a broader survey of first-principles phonon transport and the evolution from three-phonon to four-phonon treatments.

However, for \emph{metals} at $T\!\sim\!300$~K, phonon lifetimes are primarily limited by e-ph damping, and the dominant phonon mean free paths are typically below a few–tens of nanometers, so boundary/size effects are relevant (see, e.g., \cite{tong2019comprehensive}). Within this metallic regime, four-phonon scattering is \emph{often} a secondary correction to the e-ph limited lifetimes at room temperature, but it can become non-negligible in stiffer intermetallics, at elevated temperatures, or when optical branches play a larger role. When warranted, its inclusion is straightforward via Eq.~(\ref{eq:tau_total_3ph_4ph}) by adding $\tau_{\lambda}^{\mathrm{ph\text{--}ph(4)}}$ to the total scattering rate. \footnote{In Section~3 (metals, $T\!\sim\!300$~K) we model lattice thermal transport using ph-ph scattering limited to cubic (three-phonon) processes together with explicit e-ph damping; four-phonon (quartic) scattering is not included in the presented results.}
\subsection{Electron–phonon scattering}
For a given phonon mode $\lambda=(\mathbf q,\nu)$, Fermi’s Golden Rule gives the probability per unit time for an electron to scatter by emitting or absorbing that phonon. The inverse phonon lifetime reads
\begin{equation}\label{eq:tau_eph}
(\tau^{\mathrm{e\text{-}ph}}_{\lambda})^{-1}
  =\frac{2\pi}{\hbar}
   \sum_{i j \mathbf k}
   \lvert g^{\lambda}_{j,\mathbf{k+q};\,i,\mathbf k}\rvert^{2}
   \bigl(f_{i\mathbf k}-f_{j\mathbf{k+q}}\bigr)
   \delta\!\bigl(\epsilon_{i\mathbf k}-\epsilon_{j\mathbf{k+q}}+\hbar\omega_\lambda\bigr),
\end{equation}
where the summation spans all initial and final electronic bands and wave vectors. The difference of Fermi–Dirac factors enforces Pauli blocking, and the Dirac delta conserves energy while allowing for phonon absorption or emission.  This expression shows that a large electronic density of states around the Fermi level, sizeable matrix elements, or high phonon frequencies all shorten $\tau^{\mathrm{e\text{-}ph}}_{\lambda}$ and thus diminish lattice heat transport.

The coupling amplitude that mediates these transitions is
\begin{equation}\label{eq:g_eph}
g^{\lambda}_{j,\mathbf{k+q};\,i,\mathbf k}
   =\sqrt{\frac{\hbar}{2\omega_\lambda}}\;
    \matrixel{\psi_{j\mathbf{k+q}}}{\partial_{\lambda}U}{\psi_{i\mathbf k}},
\end{equation}
where $\psi_{i\mathbf k}$ is the Bloch wavefunction of band~$i$ and $\partial_{\lambda}U$ is the derivative of the self-consistent crystal potential with respect to the normalized lattice displacement of mode $\lambda$. The prefactor $\sqrt{\hbar/2\omega_\lambda}$ fixes the single-quantum normalisation. Equation~\eqref{eq:tau_eph} therefore quantifies how efficiently electrons damp individual phonon modes and set the electron-limited contribution to the phonon thermal conductivity.

\subsection{Klemens–Williams analytic limit}
Treating the conduction band as a free electron gas that interacts with long-wavelength acoustic phonons through a deformation potential, Klemens and Williams obtained a closed-form e–ph lifetime

\begin{equation}\label{eq:tau_KW}
  \bigl(\tau^{\mathrm{e\text{-}ph}}\bigr)^{-1}
  =\frac{\pi}{3}\,
   \frac{v_{g}}{v_{F}}\,
   \frac{n_e D_{\mathrm{e\text{-}ph}}^{2}\,\omega}
        {\mu\, E_{\mathrm F}},
\end{equation}

where \(v_{g}\) is the acoustic phonon group velocity, \(v_{F}\) the electronic Fermi velocity, \(n_{e}\) the conduction-electron density, \(D_{\mathrm{e\text{-}ph}}\) a deformation-potential-like constant, $\mu$ the \emph{shear modulus} (Pa), and $E_F$ the Fermi energy. 
This form is the Klemens--Williams long-wavelength limit for electron damping of acoustic modes and is \emph{linear} in $\omega$. While the prefactor captures how stronger e–ph coupling, higher carrier density, or a smaller Fermi velocity accelerate phonon damping.

Insertion of the e–ph lifetime into a Debye model for the lattice conductivity yields
\begin{equation}
\kappa_{\mathrm{ph}}^{\mathrm{(ph\text{-}ph+e\text{-}ph)}}
  =\kappa_{\mathrm{ph}}^{\mathrm{(ph\text{-}ph)}}
    \Bigl[1-\frac{\omega_{i}}{\omega_{D}}
           \ln\!\bigl(\tfrac{\omega_{D}}{\omega_{i}}+1\bigr)\Bigr], 
\qquad
\omega_{i}= \omega_{D}\,
            \frac{v_{g}}{v_{F}}\,
            \frac{\pi n_{e}}{3B\gamma_{G}^{2}}\,
            \frac{D_{\mathrm{e\text{-}ph}}^{2}}{k_{\mathrm B}T\,E_{\mathrm F}},
\end{equation}
where \(\omega_{D}\) is the Debye cut-off, \(B\) the bulk modulus, and \(\gamma_{G}\) the Grüneisen parameter.  The ratio \(\omega_{i}/\omega_{D}\) acts as a compact figure of merit: stronger deformation potentials, higher carrier densities, or lower Fermi velocities \(v_{F}\) increase \(\omega_{i}\), amplifying the logarithmic term and more strongly suppressing \(\kappa_{\mathrm{ph}}\). In weakly coupled (noble) metals, by contrast, $\omega_{i}$ stays small, so the lattice contribution is only marginally reduced.

\subsection{Electron transport under e–ph control}
While the preceding subsections quantify how electrons limit
\emph{phonon} heat flow, the same matrix elements govern 
\emph{electronic} transport.  Starting from the linearised Boltzmann
equation and Onsager reciprocity, the electrical conductivity tensor is
\begin{equation}
  \sigma^{\alpha\beta} =
  -\frac{e^{2}n_{s}}{V}
   \sum_{i\mathbf k}
     \frac{\partial f_{i\mathbf k}}{\partial\epsilon}\,
     v^{\alpha}_{i\mathbf k}\,v^{\beta}_{i\mathbf k}\,
     \tau_{i\mathbf k},
\end{equation}
where $\sigma^{\alpha \beta}$ are the Cartesian components of the electrical conductivity tensor, $\tau_{i\mathbf k}$ is the lifetime of the electronic states, $n_s$ is the spin degeneracy factor, and $V$ is the normalisation volume of the crystal used in the Brillouin-zone sum. Unless otherwise stated, the Brillouin-zone sums are over all bands/valleys and the explicit spin factor $n_s$ accounts for spin degeneracy only (no double counting).
The derivative ${\partial f_{i\mathbf k}}/{\partial\epsilon}$ strongly peaks at the Fermi energy, so only states near $E_{\mathrm F}$ contribute. The electronic part of the thermal conductivity follows from Onsager reciprocity
\begin{equation}
  \kappa_{\mathrm e}=K-T\,\sigma\,S^{2},
  \label{eq:ke}
\end{equation}
with \(K\) the open-circuit heat–current coefficient and \(S\) the Seebeck coefficient. Both transport coefficients are therefore governed by the same microscopic lifetime \(\tau_{i\mathbf k}\), which can be written as
\begin{equation}
    \frac{1}{\tau_{i\mathbf k}} = \frac{1}{\tau_{i\mathbf k}^{\mathrm{e\text{-}e}}} + \frac{1}{\tau_{i\mathbf k}^{\mathrm{e\text{-}ph}}} + \cdots
\end{equation}
In metals where the electron lifetime is dominated by e–ph interactions,
\begin{align}
  \bigl(\tau_{i\mathbf k}^{\mathrm{e\text{-}ph}}\bigr)^{-1} &=
  \frac{2\pi}{\hbar}\!
  \sum_{j\lambda}
    \bigl|g^{\lambda}_{j,\mathbf{k+q};\,i,\mathbf k}\bigr|^{2}
    \Bigl[
      (n_{\lambda}+f_{j\mathbf{k+q}})\,
      \delta(\epsilon_{i\mathbf k}+\hbar\omega_{\lambda}
             -\epsilon_{j\mathbf{k+q}})
      \nonumber\\[-2pt]
    &\hspace{3.8cm}
     +(n_{\lambda}+1-f_{j\mathbf{k+q}})\,
      \delta(\epsilon_{i\mathbf k}-\hbar\omega_{\lambda}
             -\epsilon_{j\mathbf{k+q}})
    \Bigr],
\end{align}
with $n_{\lambda}$ the Bose occupation of phonon mode $\lambda$.
Larger $\lvert g\rvert^{2}$, a denser electronic DOS near $E_{\mathrm F}$, or a higher thermal phonon population all shorten $\tau_{i\mathbf k}$,
thereby reducing both $\sigma$ and $\kappa_{\mathrm e}$\cite{Ziman1960,Grimvall1981,Allen1975_PRB}.
Ziman’s Bloch–Grüneisen analysis\cite{Ziman1960} and Grimvall’s
self-energy formalism\cite{Grimvall1981} remain standard references and analytical benchmarks for low-temperature power laws and the separation of momentum- and energy-relaxation times. Allen’s treatment of the Eliashberg
spectral function established quantitative rules for resistivity
slopes and Lorenz-number renormalisation in strongly coupled metals\cite{Allen1975_PRB,Allen1987_RMP}.

Density-functional perturbation theory combined with Wannier interpolation marked the first-principles advance and delivers state-resolved $g^{\lambda}_{j,\mathbf{k+q};\,i,\mathbf k}$ on ultradense grids, enabling parameter-free predictions of $\sigma(T)$ and $\kappa_{\mathrm e}(T)$ for transition metals\cite{Giustino2017_RMP,tong2019comprehensive}. These calculations indicate that e–ph scattering alone can reproduce the measured resistivity of Cu, Ag and Au in good agreement with experiment once vertex corrections are included\cite{Mustafa2016_PRB}.

Furthermore, numerical solvers and codes such as \textsc{BoltzTraP2}\cite{Madsen2018_CPC} and \textsc{EPW}\cite{Ponce2016_CPC} implement the linearised Boltzmann equation with onboard e–ph matrix elements, while \textsc{ShengBTE} extensions have begun to treat the coupled electron and
phonon problems on equal footing\cite{Li2020_PRL}.
Together these analytical and computational tools provide a seamless bridge from the long-established Bloch–Grüneisen picture to modern \emph{ab initio}, mode-resolved transport in metals.

\subsection{Key damping parameters}
For a given phonon mode $\lambda=(\mathbf q,\nu)$, Eq.~\eqref{eq:tau_eph} shows that the electron–phonon damping rate is controlled by three microscopic ingredients.

\textbf{(i) Electronic phase space.}
The joint availability of initial–final electronic states near $E_{\mathrm F}$ scales with the Fermi-level density of states $N(E_{\mathrm F})$ (and the detailed band velocities). A larger $N(E_{\mathrm F})$ increases the joint DOS, shortens phonon lifetimes, and lowers $\kappa_{\mathrm{ph}}$; $d$-band transition metals typify this high-DOS regime.

\textbf{(ii) Energy window set by the phonon frequency.}
The $\delta$-function in Eq.~\eqref{eq:tau_eph} restricts scattering to an energy shell of width $\hbar\omega_\lambda$. In simple three-dimensional (nearly free-electron) metals the joint DOS grows roughly linearly with $\omega_\lambda$, so high-frequency modes experience stronger damping than long-wavelength acoustics. (In monatomic fcc metals such as Cu/Ag/Au there are no optical branches; the “high-$\omega$” remark applies to multi-atom metals/intermetallics.)

\textbf{(iii) Matrix-element strength.}
The coupling $\lvert g^{\lambda}_{j,\mathbf{k+q};\,i,\mathbf k}\rvert^2$ measures the sensitivity of the self-consistent potential to atomic displacements and reflects bonding character and deformation potentials; long-range polar (Fröhlich-like) contributions are strongly screened in good metals but can matter in polar intermetallics or at low carrier density.

In combination, large $N(E_{\mathrm F})$, higher $\omega_\lambda$, and stronger $\lvert g_\lambda\rvert^2$ conspire to shorten phonon lifetimes and suppress the lattice contribution to heat flow. This explains why $d$-band transition metals and many ordered intermetallics can lose up to $\sim$40\% of $\kappa_{\mathrm{ph}}$ once e–ph scattering is included, whereas noble metals—with \emph{strong} screening and low $N(E_{\mathrm F})$—are only weakly affected.

\subsection{Dimensionality, screening, and electron–electron corrections}

\textbf{Confinement effects.}
Femtosecond pump–probe measurements on Au and Ag nanoparticles show that electron cooling times decrease markedly with size—from $\sim 850$\,fs at diameters $\sim 25$–30\,nm to $\sim 500$\,fs at $\sim 3$\,nm—reflecting enhanced surface scattering and modified e–ph coupling in confined geometries\cite{Sun1994_PRB,DelFatti2000_ChemPhys,Hartland2011_ChemRev}.  
Because the dominant heat‐carrying phonons in metals already have short mean-free paths, additional boundary scattering in thin films, nanowires, and grain-refined materials further suppresses the lattice contribution $\kappa_{\mathrm{ph}}$\cite{Cahill2003_JAP,Majumdar1993_JHT}. This can be seen in Fig. 1(b). 

\smallskip
\textbf{Electrostatic screening.}
Both e–ph and e–e rates depend on the screened Coulomb interaction, well approximated in metals by a Yukawa (Thomas–Fermi/Lindhard) potential,
\begin{equation}
V(r)=\frac{e^{2}}{4\pi\epsilon_{0}}\frac{e^{-k_{s} r}}{r},
\end{equation}
where the screening wave vector $k_{s}$ may be taken as the Thomas–Fermi value $k_{\mathrm{TF}}$ or, more generally, as the static/dynamic Lindhard function $k_{s}(q,\omega)$\cite{AshcroftMermin,kazemian2023dynamic}.  
Stronger (“hard”) screening (large $k_{s}$) reduces the range of long-range fields and weakens Fröhlich-like couplings; weaker (“soft”) screening has the opposite effect\cite{Ziman1960,Giustino2017_RMP}.

\smallskip
\textbf{Electron–electron corrections.}
For transport coefficients, the e–e contribution is sensitive to how screening is treated. Partial-wave (phase-shift) solutions of the e–e scattering problem with a consistently screened potential show that first-Born estimates overstate the e–e thermal resistivity by roughly a factor of two; using the same $k_{s}$ in both treatments restores agreement and avoids the larger ($\sim\!5\times$) discrepancies quoted in older literature\cite{QuinnFerrell1958,Gurzhi1968_Usp}.  
Accurate accounting of e–e scattering is therefore essential when deconvolving $\kappa_{\mathrm{e}}$ and $\kappa_{\mathrm{ph}}$ in nanoscale metals and in ultrafast pump–probe analyses\cite{Hartland2011_ChemRev,Cahill2003_JAP}.

\begin{table}[h]
  \centering
  \caption{Typical room-temperature lattice-conductivity ranges 
           and the dominant reasons for their magnitude.*
          \label{tab:kappa_ranges}}
  \renewcommand{\arraystretch}{1.2}
  \begin{tabular}{@{}lccccl@{}}
    \toprule
    Metal class    
      & $\kappa_{\mathrm{ph}}^{\text{(ph-ph)}}$
      & $\kappa_{\mathrm{ph}}^{\text{(ph-ph+e-ph)}}$
      & $\kappa_{\mathrm{ph}}/\kappa_{\text{tot}}$
      & Main limiting factor(s)   
      \\ \midrule
    Noble (Cu, Ag, Au)           
      & 3–10  
      & 2–6   
      & $<10\%$  
      & Low $N(E_{\mathrm F})$, weak EPC,
      \\
      &
      &
      &
      & \textbf{strong} screening
      \\[4pt]
    Transition metals 
      & 5–30  
      & 3–18  
      & 10–40\% 
      & Large $N(E_{\mathrm F})$, high $\omega_{\lambda}$,
       \\
    (Ni, Co, Pt)
      &
      &
      &
      & strong deformation-potential coupling
       \\[4pt]
    Ordered intermetallics        
      & similar to parent     
      & significantly lower       
      & variable                
      & Optical-branch EPC dominates  
       \\
    CuAu, Cu$_3$Au
      & phases
      &
      &
      & (multi-atom bases)
       \\
    \bottomrule
  \end{tabular}
  \vspace{6pt}
  \raggedright
  \footnotesize
  *Ranges are indicative values in W\,m$^{-1}$\,K$^{-1}$ at $T\!\approx\!300$ K.  
  $\kappa_{\mathrm{ph}}^{\text{(ph-ph)}}$: only phonon–phonon scattering;  
  $\kappa_{\mathrm{ph}}^{\text{(ph-ph+e-ph)}}$: phonon–phonon \emph{plus} electron–phonon scattering.  
  EPC $\equiv$ electron–phonon coupling; $N(E_{\mathrm F})$ $\equiv$ electronic density of states at the Fermi level;  
  $\omega_{\lambda}$ $\equiv$ phonon frequency.  
\end{table}

The table compares typical room-temperature lattice thermal conductivities for three metal families, first assuming ph–ph scattering alone,
$\kappa_{\mathrm{ph}}^{(\mathrm{ph\text{-}ph})}$, and then after e–ph damping is included, $\kappa_{\mathrm{ph}}^{(\mathrm{ph\text{-}ph+e\text{-}ph})}$.
In noble metals such as Cu, Ag, and Au a low electronic density of states together with \emph{strong} screening reduce the lattice channel only modestly, from
$\sim$3--10 to $\sim$2--6\,W\,m$^{-1}$\,K$^{-1}$, so phonons carry
$<10\,\%$ of the total heat flow.
Transition metals (Ni, Co, Pt) possess a large $N(E_{\mathrm F})$, stiffer phonons, and strong deformation-potential couplings; e–ph scattering therefore cuts the lattice conductivity from $\sim$5--30 to $\sim$3--18\,W\,m$^{-1}$\,K$^{-1}$, leaving phonons responsible for 10--40\,\% of overall conduction.
Ordered intermetallics such as CuAu or Cu$_3$Au behave similarly to their parent elements but can experience additional damping because optical branches couple strongly to electrons in their multi-atom bases. Hence the combined magnitude of $N(E_{\mathrm F})$, phonon frequency, and the e–ph matrix element dictates how severely the lattice channel is suppressed, ranging from almost negligible in noble metals to quantitatively significant in $d$-band and intermetallic systems.

\begin{figure}[h]
\centering
\includegraphics[width=0.95\columnwidth]{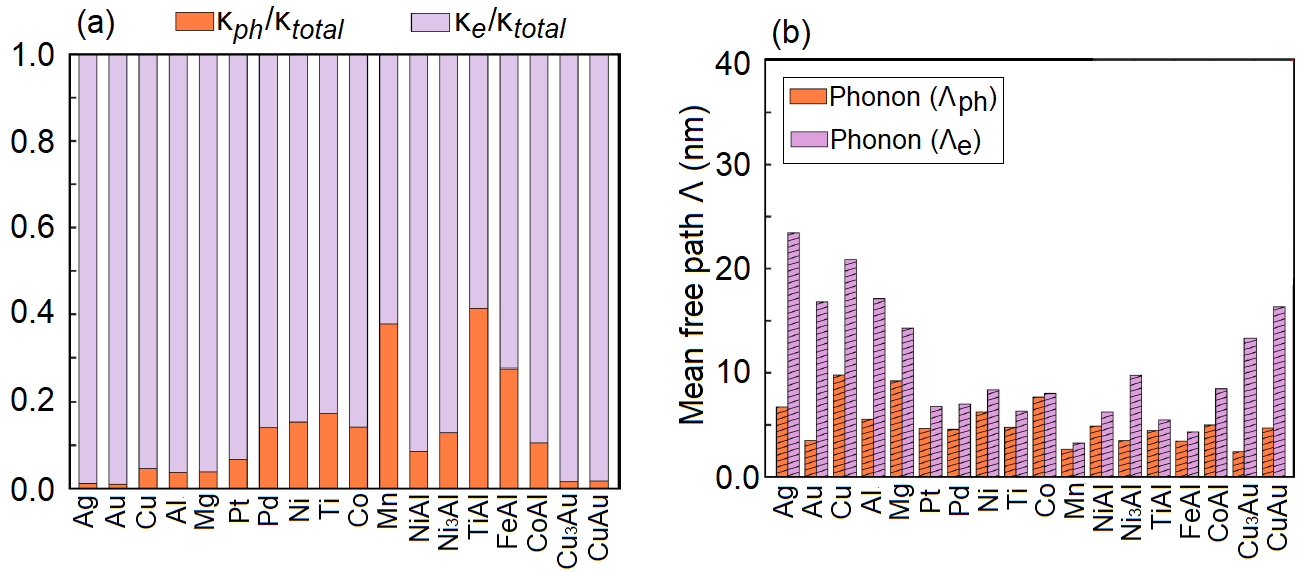}
\caption{(a) Phonon vs. electron share of $\kappa_{total}$ at 300 K: Phonons contribute $<10\%$ of $\kappa_{\mathrm{tot}}$ in noble/alkali/NIC metals but $10$–$40\%$ in transition/TIC metals (even though $\kappa_{\mathrm{ph}} \approx 3$–$15\ \mathrm{W\,m^{-1}\,K^{-1}}$). Higher $\sigma$ in nobles and stronger e-ph coupling in transitions explain the contrast. (b) Mean free paths at 300 K: Average mean free paths (MFPs) at $50\%$ accumulation of thermal conductivity show phonons $\lesssim 10\ \mathrm{nm}$ for all 18 metals, while electrons are $\sim 5$–$25\ \mathrm{nm}$; electron MFPs generally exceed phonon MFPs, implying a stronger size effect in $\kappa_{\mathrm e}$ for metal nanostructures.
Figures taken from \emph{Tong et al.}\cite{tong2019comprehensive}}
\label{Figure1}
\end{figure}

\subsection{Summary and open challenges}

Electron–phonon scattering, not ph–ph interactions, is the principal bottleneck for lattice heat transport in most bulk metals. Modern first-principles e–ph frameworks now deliver mode-by-mode values for the phonon conductivity $\kappa_{\mathrm{ph}}$ and the electronic conductivity 
$\kappa_{\mathrm e}$, replacing semi-empirical Debye–Klemens models that can miss the mark by more than a factor of three. Building on these advances, three avenues stand out for future work:

\textbf{Extend e–ph coupling data to new regimes,}
Current databases focus on non-magnetic, near-equilibrium metals at 300 K. Including finite-temperature magnetism, strong spin–orbit coupling, and ultrafast laser-excited nonequilibrium states will make predictions relevant to magnetic memory devices, spintronics, and pump–probe experiments.
  
\textbf{Benchmark against experiment,}
Systematic comparisons with Lorenz-number measurements, time-domain thermoreflectance, and femtosecond electron diffraction will validate (or refine) first-principles lifetimes, ensuring that theoretical gains translate into quantitative accuracy.
        
\textbf{Exploit nanoscale size effects,}
Phonon mean-free-path spectra show that, once limited by e–ph coupling, the dominant heat-carrying phonons are scattered after travelling only a few to a few tens of nanometres. Tailoring layer thicknesses, grain sizes, or nanowire diameters offers a practical route to engineer heat flow in nano-interconnects, metallic superlattices, and thermoelectric barriers.

In short, a concerted push to widen e–ph datasets, anchor them to precise measurements, and integrate size-dependent design rules will turn today’s qualitative understanding into predictive control of thermal transport across the metallic landscape.

\section{Electron–phonon–limited lattice heat transport in 
            semiconductors}

\subsection{Ab initio foundation}

We use the density functional theory (DFT) to obtain the
electronic eigenenergies $\epsilon_{n\mathbf k}$ and wave-functions
$\psi_{n\mathbf k}$, followed by density-functional perturbation theory (DFPT)
for phonon frequencies $\omega_{\mathbf q\nu}$ and eigendisplacements
$e_{\nu\mathbf q}$. Within the DFT/DFPT formalism\cite{baroni2001phonons}, the linear, self-consistent
change in the Kohn--Sham potential associated with a phonon mode $(\nu,\mathbf q)$
can be written as a mass–weighted sum of Cartesian derivatives:
\begin{equation}
  \Delta_{\nu\mathbf q} V(\mathbf r)
  = \sum_{\kappa\alpha}\frac{e^{\nu\mathbf q}_{\kappa\alpha}}{\sqrt{M_\kappa}}\,
    \partial_{\mathbf q,\kappa\alpha} V(\mathbf r),
  \qquad
  \partial_{\mathbf q,\kappa\alpha} V(\mathbf r)
  = \sum_{\mathbf R_p} e^{i\mathbf q\cdot\mathbf R_p}\,
    \frac{\partial V(\mathbf r)}{\partial R_{p\kappa\alpha}},
\end{equation}
where $e^{\nu\mathbf q}_{\kappa\alpha}$ are the phonon eigenvectors and $M_\kappa$ the
ionic masses. The corresponding electron--phonon (e–ph) matrix element is\cite{Giustino2017_RMP}
\begin{equation}
  g_{mn}^{\nu}(\mathbf k,\mathbf q)
  = \sqrt{\frac{\hbar}{2\omega_{\mathbf q\nu}}}\;
    \big\langle \psi_{m,\mathbf{k+q}}\big|\,\Delta_{\nu\mathbf q} V\,\big|\psi_{n,\mathbf k}\big\rangle,
  \label{eq:g-def}
\end{equation}
with $\psi_{n\mathbf k}$ being the
Bloch states. 

\begin{figure}[h]
\centering
\includegraphics[width=0.9\columnwidth]{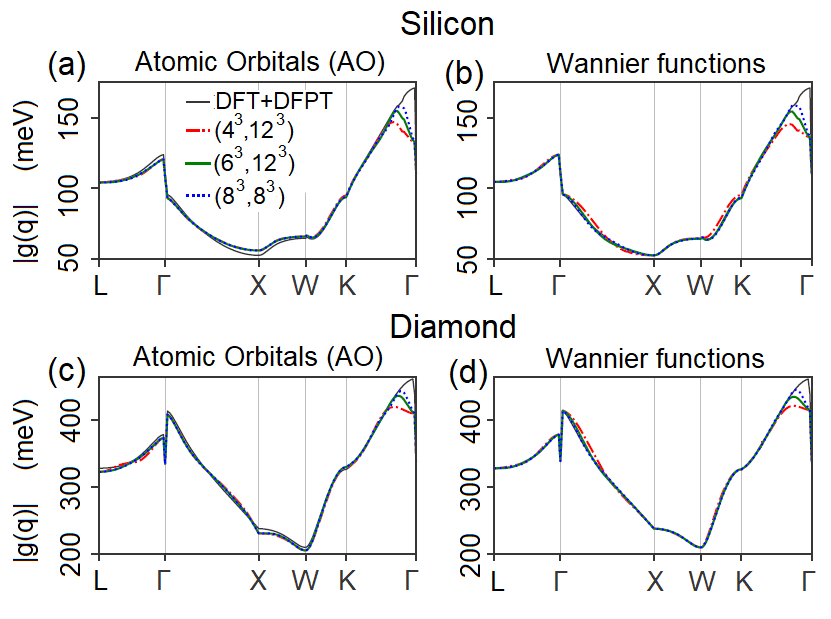}
\caption{Interpolated e-ph matrix elements for (a), (b) silicon and (c), (d) diamond, for different coarse grids. We observe the accuracy of e–ph matrix-element interpolation. Both AO and WF schemes reproduce direct-DFPT \(|g_{mn}^{\nu}|\) along high-symmetry lines within numerical precision, validating the coarse\(\to\)dense interpolation.
Figures taken from \emph{Agapito et al.}\cite{Agapito2018_PRB}}
\label{Figure2}
\end{figure}

In polar crystals, recovering the correct $q\!\to\!0$ behaviour
requires adding the long-range Fr\"ohlich contribution analytically to the
short-range DFPT part\cite{Giustino2017_RMP,Verdi2015_PRL}. Direct evaluation of Eq.~\eqref{eq:g-def} on ultra-dense $(\mathbf k,\mathbf q)$ grids
is prohibitively expensive. Modern workflows therefore interpolate the \emph{same} matrix elements from coarse DFPT meshes to dense grids using maximally localised Wannier functions (e.g.\ EPW)\cite{Ponce2016_CPC} or atomic-orbital
interpolation\cite{Agapito2018_PRB}. This is shown in Fig. 2. Given $g_{mn}^{\nu}$, the electron-limited phonon lifetime follows from Fermi’s
golden rule\cite{Giustino2017_RMP}:
\begin{equation}
\bigl(\tau^{\mathrm{e\text{-}ph}}_{\nu\mathbf q}\bigr)^{-1}
= \frac{2\pi}{\hbar}
  \sum_{mn\mathbf k}
    \bigl|g^{\nu}_{mn}(\mathbf k,\mathbf q)\bigr|^{2}
    \bigl[f_{n\mathbf k}-f_{m,\mathbf{k+q}}\bigr]\,
    \delta\!\bigl(\epsilon_{n\mathbf k}-\epsilon_{m,\mathbf{k+q}}+\hbar\omega_{\mathbf q\nu}\bigr),
\end{equation}
where the Fermi factors enforce Pauli blocking and the Dirac delta imposes energy conservation.
These mode-resolved lifetimes enter the single-mode RTA for the lattice thermal
conductivity,
\begin{equation}
  \kappa_{\mathrm{ph}}^{\alpha\beta}
  = \sum_{\nu\mathbf q} c_{v,\nu\mathbf q}\,
    v^{\alpha}_{\nu\mathbf q}\, v^{\beta}_{\nu\mathbf q}\,
    \tau_{\nu\mathbf q},
\end{equation}
with $c_{v,\nu\mathbf q}$ the modal heat capacity and
$\mathbf v_{\nu\mathbf q}=\partial\omega_{\mathbf q\nu}/\partial\mathbf q$
the group velocity\cite{Giustino2017_RMP}. All branch-resolved trends below stem
from inserting the \emph{ab initio} lifetimes into this expression.


\subsection{The role of different phonon branches}

In covalent semiconductors such as Si and Ge, long-wavelength acoustic (LA/TA) modes couple to carriers through the (short-range) deformation-potential interaction. The corresponding e–ph matrix elements vanish at $q\!\to\!0$ and increase with $|q|$; together with the linear acoustic dispersion $\omega=v_s q$, this makes carrier scattering by the very lowest-frequency LA/TA modes weak. As a result, these modes retain long mean free paths and dominate the lattice thermal conductivity at room temperature.\cite{Agapito2018_PRB,Li2015_PRB,Broido2007_APL}

In polar III–V semiconductors the situation changes: the long-range Fr\"ohlich field couples carriers strongly to longitudinal-optical (LO) phonons, and the mode dependence of the e–ph matrix elements becomes essential.\cite{Verdi2015_PRL,Sjakste2007_PRL,Sjakste2015_PRB} When electrons and phonons are treated self-consistently in coupled BTE frameworks, momentum/energy exchange mediated by e–ph scattering can noticeably reduce the phonon thermal conductivity relative to a phonon–phonon-only model, especially at finite carrier densities.\cite{Protik2022_npj,Protik2020_PRB}

Two-dimensional semiconductors add a quadratic flexural (ZA) branch. DFPT together with anharmonic renormalization shows that, at room temperature and for carrier densities $\lesssim 10^{12}\,\mathrm{cm^{-2}}$, ZA scattering contributes less than $\sim\!30\%$ of the total resistive e–ph rate; in-plane acoustic modes thus dominate both carrier mobility and lattice heat conduction under typical conditions.\cite{Rudenko2019_PRB}

\subsection{Carrier density and screening}

Heavy doping reshapes electron--phonon physics in two synergistic ways.
First, each additional carrier enlarges the Fermi sphere
($k_{F}=(3\pi^{2}n)^{1/3}$), widening the phase space for intravalley acoustic deformation-potential scattering in the degenerate, quasi-elastic limit. A convenient form is
\begin{equation}
\frac{1}{\tau^{\mathrm{e\text{-}ph}}_{\mathrm{ac}}}
      \simeq C\,
      \frac{D_{\mathrm{ac}}^{2}\,k_{\mathrm B}T}{\rho v_{s}^{2}}\,
      \frac{m^{\!*}k_{F}}{\hbar^{3}v_{s}},
\label{eq:tau_ac_deg}
\end{equation}
where \(D_{\mathrm{ac}}\) is the acoustic deformation potential, \(\rho\) the mass density, \(v_{s}\) the sound velocity, and \(C\sim\mathcal{O}(1)\) is an angular factor that depends weakly on band anisotropy and degeneracy\cite{Ziman1960,Li2015_PRB}. 

Second, the dense electron gas strengthens electrostatic screening. Rather than equating the screening vector to \(k_F\), one should use a Thomas--Fermi/Lindhard wavevector \(k_{s}\) that \emph{increases} with \(n\) (via \(k_F\)) but is not equal to it; transport coefficients are notably sensitive to the choice of \(k_s\)\cite{Kukkonen1973_PRB}. In \emph{ab initio} BTE studies of $n$-type Si, once
\(n\!\gtrsim\!10^{19}\,\mathrm{cm^{-3}}\) the calculated electron-limited damping of low-frequency LA/TA modes grows rapidly and the resulting suppression of \(\kappa_{\mathrm ph}\) tracks time-domain thermoreflectance benchmarks\cite{Li2015_PRB,Protik2022_elphbolt}.

In polar III--V materials such as GaN, the long-range Fr\"ohlich interaction couples carriers most strongly to LO phonons. A standard screened form is
\begin{equation}
\bigl|g_{\mathrm F}(q)\bigr|^{2}=
  \frac{e^{2}\hbar\omega_{\mathrm{LO}}}{2\epsilon_{0}V}
  \!\left(\frac{1}{\epsilon_{\infty}}
        -\frac{1}{\epsilon_{s}}\right)\!
  \frac{1}{q^{2}+k_{s}^{2}},
\end{equation}
so increasing \(k_s\) (heavier doping) softens the long-range field. Nonetheless, room-temperature energy relaxation and phonon damping remain LO-dominated over wide doping ranges in nitrides; screening reduces the rate but does not eliminate the LO bottleneck\cite{Jhalani2020_PRL,Zhang2001_JPCM}. In short, heavy doping in covalent Si throttles heat flow primarily by empowering low-frequency acoustics, whereas in polar GaN it leaves the LO channel as the primary sink—highlighting how bonding character (deformation potential vs.\ Fr\"ohlich) and carrier screening jointly decide which branch ultimately limits the lattice thermal conductivity.

\subsection{Dimensionality and dielectric environment}

Reducing a semiconductor to monolayer or ultrathin‐film thickness fundamentally modifies the microscopic ingredients of the e–ph interaction. Confinement quantises the out‐of‐plane phonon wave vector, replacing the 3D continuum $q_{z}$ by discrete values $q_{z}=n\pi/L$ ($n=0,1,\dots$), while the in‐plane component $\mathbf q_{\parallel}$ remains continuous down to $\mathbf q_{\parallel}\!\to\!0$. The \emph{in‐plane} acoustic branch with $n=0$ remains gapless with $\omega=v_{s} q_{\parallel}$; by contrast, the first thickness-quantised subband ($n=1$) acquires a gaplike term,
\begin{equation}
\omega_{\mathrm{LA},1}(\mathbf q_{\parallel})=
  v_{s}\sqrt{q_{\parallel}^{2}+(\pi/L)^{2}},
\end{equation}
so long‐wavelength 3D channels that require finite $q_z$ below $\pi/L$ are kinematically excluded even though arbitrarily small in‐plane momenta are allowed. A simple way to encode this thickness cutoff in the out‐of‐plane phase space is
\begin{equation}
N_{\text{film}}(q_{z})=\Theta\!\bigl(|q_{z}|-q_{\min}\bigr),
\qquad
q_{\min}\approx\frac{\pi}{L}.
\end{equation}

This $\Theta$-filter is a heuristic to encode the loss of continuous $q_z$ phase space; in real films the spectrum consists of discrete subbands (Lamb modes) rather than a hard cutoff.
At the same time, freestanding monolayers host a quadratic flexural (ZA) branch with $\omega_{\mathrm{ZA}}=Aq^{2}$. For crystals with horizontal mirror symmetry, the leading e–ph coupling to ZA modes is symmetry‐forbidden at first-order and emerges at higher order, giving a small‐$q$ scaling $\lvert g_{\mathrm{ZA}}\rvert\!\propto\! q^{2}$; as a result, the intrinsic ZA contribution to resistive scattering is strongly suppressed compared to in‐plane acoustics unless symmetry is broken by substrates, strain, or gating.\cite{Rudenko2019_interplay} First‐principles transport studies in 2D semiconductors consistently find that in‐plane acoustic modes dominate and that the role of flexural modes is reduced to $\lesssim\!30\%$ at low carrier densities (with further suppression on substrates).\cite{Rudenko2019_interplay,Liao2015_Phosphorene}

Confinement also shortens relevant mean‐free paths compared with bulk. In monolayer MoS$_2$, \emph{electronic} mean free paths at 300 K are typically below $\sim\!9$\,nm when phonon scattering limits transport\cite{Li2015_PRB}. This has been shown in Fig. 3(a). phonon MFPs likewise contract relative to the bulk parents when boundary, substrate, and e–ph mechanisms are included, though precise values are material- and environment‐dependent.

A second confinement effect is electrostatic: the long‐range Fr\"ohlich interaction in 2D depends on an \emph{effective} out‐of‐plane permittivity set by the surrounding media,
\begin{equation}
\epsilon_{\mathrm{eff}}
       =\frac{\epsilon_{\text{sub}}+\epsilon_{\text{env}}}{2},
\qquad
g_{\mathrm F}^{\text{2D}}(q)
       =\sqrt{\frac{e^{2}\hbar\omega_{\mathrm{LO}}}
                    {2\epsilon_{0}A}}\,
         \frac{1}{\bigl(q+q_{\text{TF}}\bigr)\,\epsilon_{\mathrm{eff}}},
\end{equation}

where $q_{\text{TF}}$ is the 2D Thomas–Fermi wave vector and $A$ the monolayer area. Increasing $\epsilon_{\mathrm{eff}}$ (e.g.\ by h‐BN encapsulation) weakens the 2D Fr\"ohlich vertex and lengthens LO lifetimes; conversely, high‐$\kappa$ substrates (e.g.\ HfO$_2$) enhance polar damping. First‐principles treatments of polar e–ph coupling explicitly include the long‐range Fr\"ohlich part of the vertex and its screening,\cite{Giustino2017_RMP,Verdi2015_FRLVtx,Sjakste2015_WannierPolar} and coupled e–ph BTE solvers now propagate these interactions self‐consistently through the electron and phonon distributions under the same driving fields.\cite{Protik2022_elphbolt}

Dimensional confinement replaces the bulk's continuous out-of-plane wavevector by \emph{discrete} thickness subbands (Lamb modes), thereby removing the continuous \(q_z\) phase space. The in-plane acoustic branches remain gapless (with \(\omega \approx v_s q_\parallel\)), and, when the membrane is freestanding and mirror-symmetric, a flexural (ZA/\(A_0\)) branch survives with quadratic dispersion \(\omega \propto q_\parallel^2\) at long wavelength. Strong substrate coupling or clamped boundaries can gap, hybridize, or damp this flexural mode. The combination compresses the relevant mean‐free‐path spectra into the few–tens‐of‐nanometres window and, because $\epsilon_{\mathrm{eff}}$ is substrate‐dependent, renders the e–ph interaction—and therefore the lattice thermal conductivity—highly tunable via substrate choice, encapsulation, or electrostatic gating.


\subsection{Coupled BTE and the role of \textsc{elphbolt}}

The prevailing \emph{ab initio} workflow treats electrons and phonons in parallel:
one solves a phonon Boltzmann transport equation (BTE) for $\kappa_{\mathrm ph}$ and an electron BTE for $\sigma$ and $S$. In reality the two distribution functions are coupled: non-equilibrium electrons push phonons out of equilibrium (\emph{electron drag}), and drifting phonons pull electrons along (\emph{phonon drag}). Historically, semi-analytical drag models partially decoupled the BTEs and introduced effective terms, but this can violate the Kelvin–Onsager relationship connecting Seebeck and Peltier responses\cite{Protik2022_npj,Sondheimer1956, Herring1954, Zhou2015_PNAS, Mahan2014_JAP}.

\textsc{elphbolt} solves the \emph{fully coupled}, linearised electron–phonon BTEs from first principles. It iterates the electron and phonon collision integrals to self-consistency under the same driving fields ($\nabla T$ and $E$), and explicitly enforces the Kelvin–Onsager reciprocity at each iteration. Crucially, it uses exactly the same DFPT–Wannier electron–phonon matrix elements as single-species BTE workflows, so one can switch from the independent-channel picture to the coupled one without refitting any parameters\cite{Giustino2017_RMP,Protik2022_npj,Sjakste2015_Wannier, Verdi2015_Frohlich}.

Concrete results illustrate when coupling matters. In $n$-type Si, \textsc{elphbolt} reproduces the temperature dependence of the Seebeck coefficient across low and high doping; the phonon-drag contribution becomes dominant below $\sim\!175$~K, while at room temperature direct e–ph scattering dominates and single-BTE and coupled-BTE predictions nearly coincide. For $\kappa_{\mathrm ph}$ in Si, the \emph{electron drag} correction is small; suppression of $\kappa_{\mathrm ph}$ at low temperature is governed mainly by the phonon–electron \emph{scattering} rates rather than mutual drag. Similar “small drag on $\kappa_{\mathrm ph}$” conclusions hold for GaAs and SiC\cite{Protik2022_npj, Protik2020_GaAs, Protik2020_SiC}. This has been shown in Fig. 3(b). In strongly polar materials (e.g.\ GaN), accurate treatment of long-range polar couplings may require quadrupolar corrections; until those are included, drag conclusions should be drawn with care\cite{Protik2022_npj,Sjakste2015_Wannier,Verdi2015_Frohlich}.

In short: use coupled e–ph BTEs when phonon or electron \emph{drag} is relevant (low-to-intermediate $T$, clean samples, or structures with suppressed impurity scattering), or when enforcing Kelvin–Onsager consistency matters for quantitative thermoelectric predictions. Otherwise, single-BTE solutions (RTA or iterative) are often sufficient and substantially cheaper\cite{Protik2022_npj}.

\begin{figure}[h]
\centering
\includegraphics[width=0.95\columnwidth]{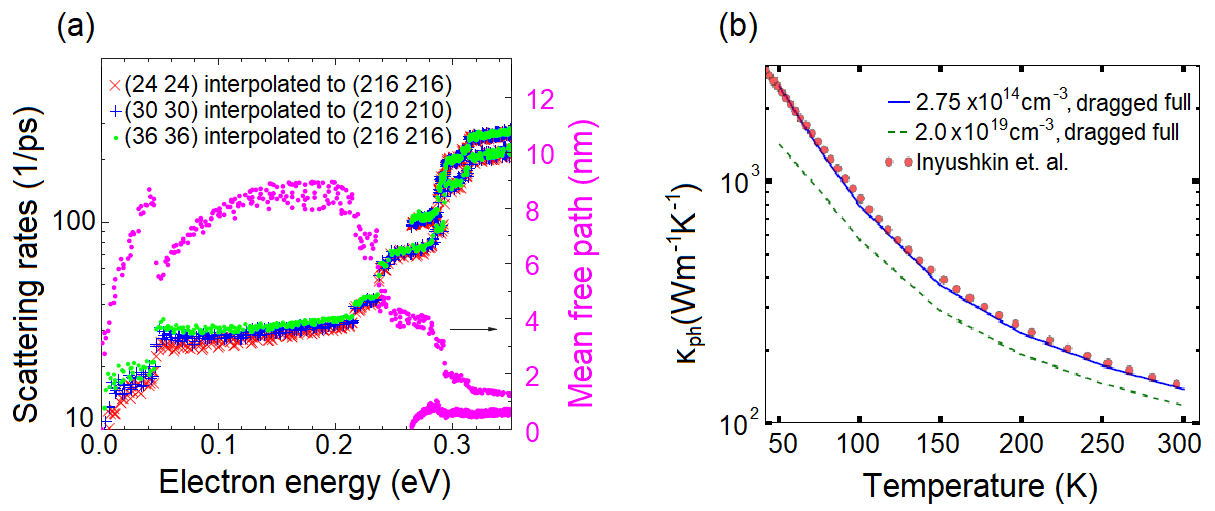}
\caption{(a) In monolayer MoS$_2$ at 300 K, phonon-limited \emph{electronic} mean free paths cluster in the few–10\,nm range (often $<\!10$\,nm), illustrating the strong size/confinement sensitivity and providing context for the 2D transport lengths discussed in this subsection. Figure taken from \emph{Li et al.}\cite{Li2015_PRB}. (b) Lattice thermal conductivity of Si vs temperature at low and high doping from the fully coupled e–ph BTE agrees with experiment \emph{Inyushkin et al.}\cite{inyushkin2004isotope} and shows stronger low-$T$ suppression from e-ph scattering. At room temperature ph–ph scattering dominates and coupled/uncoupled predictions converge. Figure taken from \emph{Protik et al.}\cite{Protik2022_npj}}
\label{Figure3}
\end{figure}

\subsubsection*{Experimental validation}

A consistent body of measurements now benchmarks the fully first-principles workflow (DFPT/Wannier e–ph vertices + BTE solvers). In \emph{boron-doped silicon}, thermal‐conductivity data on both natural and isotopically enriched crystals, together with doping-dependent measurements, exhibit the expected suppression of $\kappa_{\mathrm ph}$ as carriers are added and temperature is lowered. Fully coupled e–ph BTE calculations (\textsc{elphbolt}) reproduce the measured temperature dependence of $\kappa_{\mathrm ph}$ for $n$‐type Si across dilute and degenerate limits, and quantify why the phonon–electron channel suppresses $\kappa_{\mathrm ph}$ more strongly at low $T$ (where long-wavelength acoustics dominate) than at high $T$ (where ph–ph scattering wins)\cite{Protik2022_npj}. These comparisons are made directly against high-purity Si data and are consistent with isotope‐controlled measurements (which isolate mass-disorder effects) and with earlier doping-dependent thin-film studies\cite{Inyushkin2018_SiIsotope,Asheghi2002_SiFilms}.%
\footnote{In \textsc{elphbolt}’s Si case study, the measured points for $\kappa_{\mathrm ph}(T)$ at low and high doping (including $n\!\sim\!2\times 10^{19}\,$cm$^{-3}$) are reproduced without empirical lifetimes, and electron-drag corrections to $\kappa_{\mathrm ph}$ are found to be small; see Fig. 6 and discussion in Ref.~\cite{Protik2022_npj}.}

For \emph{polar III–V semiconductors} (e.g.\ GaAs), experiments using transient thermoreflectance and related pump–probe probes report reductions of lattice thermal conductivity compatible with LO-phonon damping at finite carrier densities. First-principles treatments that include the long-range polar (Fr\"ohlich) vertex capture these trends; coupled e–ph BTE solutions further show that, while phonon drag can dominate the Seebeck response at low $T$, the electron-drag correction to $\kappa_{\mathrm ph}$ remains small in GaAs, consistent with the mode-resolved balance of ph–e versus ph–ph scattering\cite{Protik2022_npj,Sjakste2007_PRL}.%

In 2D \emph{semiconductors}, measurements on monolayer MoS$_2$ consistently find a strong substrate/encapsulation dependence of the in-plane thermal conductivity, with reported room-temperature values for suspended or weakly coupled monolayers typically tens to $\sim\!$80\,W\,m$^{-1}$\,K$^{-1}$ and lower values on amorphous oxides\cite{Zhu2023_RSC,Taube2015_MoS2_Raman}. These trends are aligned with first-principles expectations: the dielectric environment tunes the long-range polar coupling, and the dominant in-plane acoustic branches set the heat flux; \emph{ab initio} transport in monolayer MoS$_2$ also finds short \emph{electronic} mean free paths and comparatively modest Fr\"ohlich coupling in the intrinsic limit\cite{Li2015_PRB}. Encapsulation in high-quality h-BN further improves interfacial heat dissipation and can raise the effective thermal conductance of hBN/MoS$_2$/hBN stacks\cite{Ye2021_hBNencap}. 

Taken together—covalent Si (with isotope/doping controls), polar GaAs (LO-damping trends), and substrate-tunable 2D MoS$_2$—these benchmarks show that once DFPT e–ph matrix elements are in hand, first-principles transport reproduces both magnitudes and temperature trends of $\kappa_{\mathrm ph}$ without empirical lifetimes, while clarifying where mutual e–ph drag matters and where it does not\cite{Protik2022_npj}.


\subsection{Summary and open challenges}

The integrated DFT\(\rightarrow\)DFPT\(\rightarrow\)Wannier\(\rightarrow\)BTE workflow has matured into a genuinely predictive tool: starting from the ground-state charge density, it delivers mode- and state-resolved electron–phonon matrix elements \(g_{mn}^{\nu}(\mathbf{k},\mathbf{q})\), includes the required long-/short-range partitioning for polar couplings, and feeds the resulting first-principles lifetimes into Boltzmann solvers to obtain \(\kappa_{\mathrm{ph}}\) without adjustable parameters.\cite{Giustino2017_RMP,Ponce2016_CPC,Verdi2015_PRL,Agapito2018_PRB} 
In bulk Si, GaAs, and GaN, predictions are in good agreement with time-domain thermoreflectance and isotope-controlled benchmarks; in monolayer MoS\(_2\), the same framework captures substrate/encapsulation trends by varying the external dielectric screening that modulates the long-range polar vertex.\cite{Li2015_PRB,Protik2022_npj,Jhalani2020_PRL,Inyushkin2018_SiIsotope,Asheghi2002_SiFilms,Taube2015_MoS2_Raman,Ye2021_hBNencap} 
This hierarchy already incorporates dimensionality reduction, isotope engineering, and alloy/disorder scattering, turning what was once qualitative “trend spotting’’ into an engineering-grade calculator for heat flow.\cite{Giustino2017_RMP,Ponce2016_CPC}

The next frontier is to widen this \emph{ab initio} umbrella.
\emph{First}, correlated or magnetic semiconductors call for beyond-DFT band structures (DFT\(+U\), hybrid functionals, or even \(GW\)/DMFT) so that the subsequent e–ph couplings rest on accurate quasiparticle states.\cite{Giustino2017_RMP}
\emph{Second}, van der Waals heterostructures demand a truly nonlocal description of screening: the effective Fr\"ohlich vertex in, e.g., MoS\(_2\)/h-BN/graphene stacks depends on the full dielectric profile rather than a single scalar constant, and first-principles long-range corrections must be retained explicitly.\cite{Verdi2015_PRL,Sjakste2015_Wannier}
\emph{Third}, pump–probe spectroscopy pushes materials into ultrafast, far-from-equilibrium regimes where hot carriers and nonthermal phonons coevolve; coupling the self-consistent electron– and phonon–BTE engine (\textsc{elphbolt}) to optical drive/relaxation protocols promises a parameter-free view of picosecond energy flow and drag phenomena.\cite{Hartland2011_ChemRev,Protik2022_npj}

Completing these extensions should elevate today’s steady-state agreement into a comprehensive, time-resolved roadmap for thermal management and thermoelectric optimisation in next-generation electronic, photonic, and quantum devices.

\section{2D Dirac crystals and Higher-Order e-ph interactions}

Dirac crystals are gapless systems whose low–energy carriers follow a linear (Dirac) dispersion rather than the usual parabolic one. In graphene, conduction and valence cones touch at discrete Dirac points, and carriers near these points are well described by a massless Dirac Hamiltonian with an approximately constant Fermi velocity \(v_F\).\cite{CastroNeto2009_RMP,DasSarma2011_RMP} This “relativistic” band structure underlies characteristic transport phenomena such as the unconventional (half-integer) quantum Hall effect and Klein tunnelling.\cite{CastroNeto2009_RMP,Mariani2008_PRL}

In pristine, charge-neutral graphene the Fermi “surface’’ collapses to a point and the density of states vanishes linearly at the Dirac point. Modest electrostatic gating shifts \(E_{\mathrm F}\) by tens to hundreds of meV, and—since the 2D Dirac DOS grows \(\propto |E|\)—yields large changes in carrier density.\cite{CastroNeto2009_RMP,DasSarma2011_RMP} Doping then affects electron–phonon (e–ph) physics in two ways: it enlarges the electronic phase space for scattering and strengthens screening. The net impact is mode- and \(\mathbf q\)-selective: short-range deformation-potential couplings are efficiently screened, whereas the “gauge-field” part of the in-plane acoustic coupling remains essentially unscreened; moreover, in suspended graphene the linear coupling of out-of-plane (ZA) flexural modes is symmetry-forbidden at first-order and arises in pairs (a higher-order process), while gating or substrates can break mirror symmetry and activate linear ZA coupling.\cite{SohierThesis2015,Mariani2008_PRL,Sohier2017_PRB} In practice, lattice thermal conductivity in graphene is most sensitive to support/encapsulation, disorder and strain, with electrostatic control mainly influencing electronic transport. Dielectric environment and symmetry breaking can nevertheless feed back onto phonon lifetimes through changes in long-range fields and selection rules.\cite{NikaBalandin2017_RPP}

\subsection{Electronic screening: Thomas--Fermi vs Lindhard (static/dynamic)}

In calculating the e-ph interaction, the Thomas-Fermi approximation—predicated on a large, nearly spherical Fermi surface and ultrashort screening lengths—breaks down in 2D Dirac crystals, where the Fermi “surface’’ collapses to a point at low carrier density and e-ph coupling is strongly wave–vector selective. In this regime the electronic screening length can be comparable to, or even exceed, the phonon wavelength, and the static Thomas-Fermi response fails to reproduce the sharp enhancement that occurs as the phonon wave vector approaches the \(2k_F\) nesting condition. Evaluating the full Lindhard polarisation for \emph{acoustic} phonons yields a dielectric function with a pronounced cusp at \(q=2k_F\) in the \emph{static} limit \((\omega=0)\) and a strong \emph{dynamic} enhancement (divergence in idealised limits) near \(q=2k_F\) for \(\omega\neq 0\). Behaviour entirely missed by Thomas-Fermi\cite{kazemian2023dynamic}. This replaces the oversimplified Thomas-Fermi picture with a \(q\)- and \(\omega\)-dependent screening that faithfully captures Dirac-specific e-ph physics.

A convenient form of the Lindhard dielectric function is
\begin{equation}
\varepsilon(q,\omega)=1 - V(q)\,\Pi(q,\omega),
\end{equation}
where \(V(q)\) is the \emph{bare} Coulomb interaction (e.g., \(V(q)=2\pi e^{2}/\kappa q\) in 2D, \(4\pi e^{2}/\kappa q^{2}\) in 3D) and \(\Pi(q,\omega)\) is the noninteracting density–density response (Lindhard polarisation), here evaluated on the \emph{phonon} frequency \(\omega=\omega_{\nu\mathbf{q}}\). Two hallmarks follow for \emph{acoustic} modes that TF cannot capture: (i) a \emph{cuspidal} feature at \(q=2k_F\) in the \emph{static} limit (\(\omega=0\)), and (ii) a strong \emph{dynamic} enhancement near \(q=2k_F\) for \(\omega\neq 0\) (regularised in practice by finite temperature, disorder, and intrinsic damping), precisely where phase space is largest (see Fig.~4 for a schematic comparison). These features directly enter the screened e-ph matrix elements,
\begin{equation}
g_{mn\nu}^{\mathrm{scr}}(\mathbf{k},\mathbf{q}) \;\propto\; \frac{g_{mn\nu}^{(0)}(\mathbf{k},\mathbf{q})}{\varepsilon(q,\omega_{\nu\mathbf{q}})},
\end{equation}
and thus the phonon self–energy and linewidth\cite{jishi2013feynman,kazemian2023dynamic,allen1972neutron,debernardi2002anharmonic},
\begin{equation}
\Gamma_{\nu}(\mathbf{q}) \;\propto\; \big|g_{mn\nu}^{\mathrm{scr}}(\mathbf{k},\mathbf{q})\big|^{2}\,\Im\Pi(q,\omega_{\nu\mathbf{q}}),
\end{equation}
which feed directly into \(\kappa_{\mathrm{ph}}\) and coupled e-ph transport.

\emph{Practical guidance.} 
For metals or high–density 2D systems with \(q\!\ll\!k_F\), Thomas-Fermi or \emph{static} Lindhard screening is often adequate. In low– to moderately doped 2D Dirac crystals, at least the \emph{static} Lindhard response is required to capture the \(2k_F\) cusp; whenever the phonon frequency \(\omega_{\nu\mathbf{q}}\) is relevant (acoustic branches, finite-\(T\) broadening), \emph{dynamic} Lindhard screening should be used. Finite temperature smooths the static cusp and finite lifetimes regularise the dynamic enhancement, but the strong wave-vector selectivity around \(q\!\approx\!2k_F\) remains the key qualitative difference from Thomas-Fermi. In the long-wavelength limit \(q\!\to\!0\) at fixed (large) \(k_F\), both Thomas-Fermi and Lindhard recover the familiar Thomas-Fermi-like \(1/q\) trend.

\begin{figure}[h]
\centering
\includegraphics[width=0.95\columnwidth]{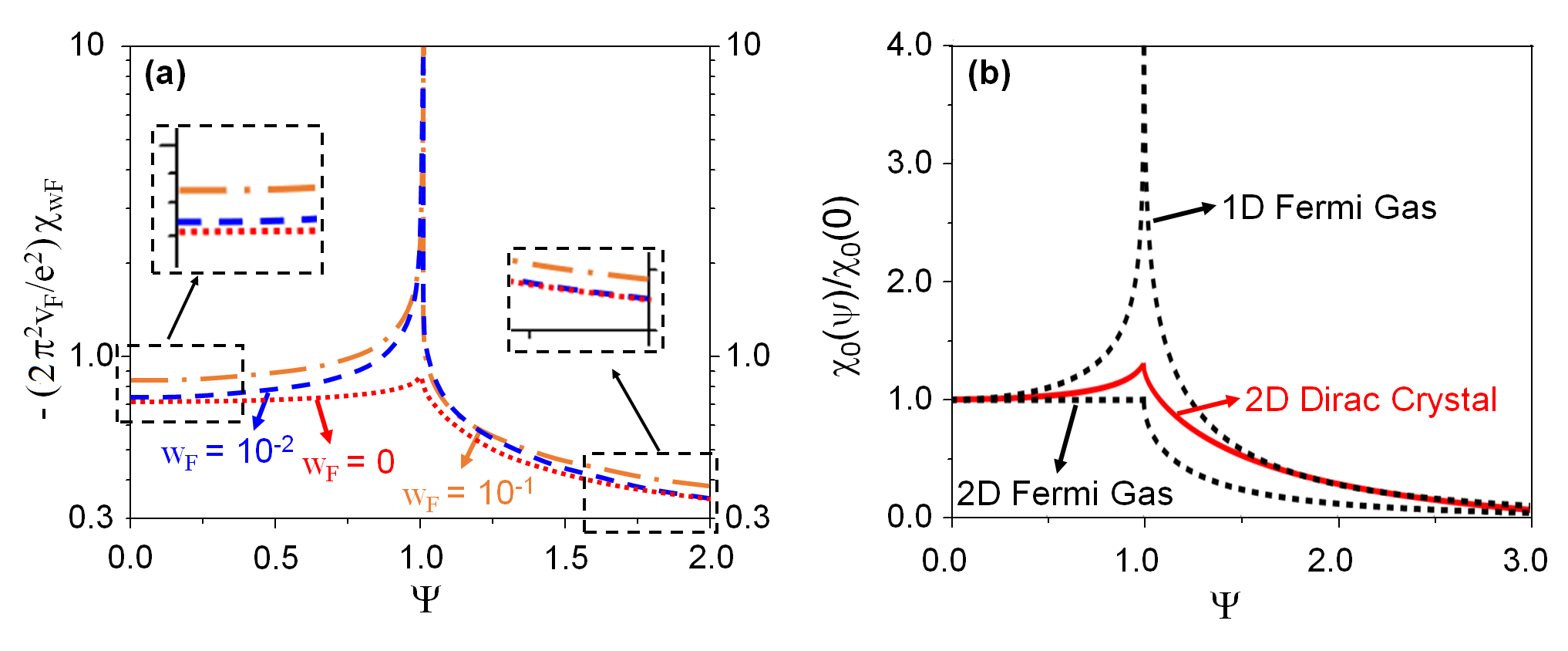}
\caption{(a) The dielectric response function $\chi$ vs $\Psi=q/2\mathrm{k}_{\mathrm{F}}\approx1$ for different 2D Dirac crystals with different $\mathrm{w_{F}} = c/\mathrm{v}_{\mathrm{F}}$ at T = 0 K. We observe that the dielectric response function diverges near the FSN at the point $\Psi\approx1$ even for small values of $\mathrm{w_{F}}\ne0$. The negligible difference in the dielectric response function of various 2D Dirac crystals with different values of $\mathrm{w_{F}}$  shows the generality of our solution for 2D Dirac crystals. (b) Comparing the dielectric response function for the static case $c=0$, designated with $\chi_{0}(\Psi)$, of a 2D Dirac crystal with that of a 1D and 2D Fermion gas at T = 0 K. We observe that $\chi_{0}(\Psi)$ of a 2D Dirac crystal does not follow a constant line like the 2D Fermi gas and it increases reaching a maximum at the FSN point, making the study of 2D Dirac crystals more intricate. Figure taken from \emph{Kazemian et al.}\cite{kazemian2023dynamic}.}
\label{Figure4}
\end{figure}

\subsection{Branch dependence of e-ph lifetimes and their impact on the thermal conductivity}

In a 2D Dirac crystal such as graphene the in–plane acoustic branches are linear,
$\omega_{\mathrm{LA/TA}}(q)=v_{s}\,q$, whereas the out–of–plane flexural branch
is quadratic,
\begin{equation}
\omega_{\mathrm{ZA}}(q)=\sqrt{\kappa_b/\rho}\;q^{2},
\end{equation}
with bending rigidity $\kappa_b$ and areal mass density $\rho$.
In two dimensions these dispersions imply that the phonon density of states
(DOS) scales as $D_{\mathrm{LA/TA}}(\omega)\!\propto\!\omega$ for linear modes
and is \emph{approximately constant} for the quadratic ZA branch
(contrary to a $\omega^{-1/2}$ scaling); this difference is central to the
low–frequency phase space in 2D.\cite{Lindsay2014_PRB,Gu2018_RMP,Nika2017_RPP}

The electron–limited phonon linewidth (inverse lifetime) for a mode $(\nu,\mathbf q)$
follows from Fermi’s golden rule:
\begin{equation}
\Gamma_{\nu\mathbf q}\equiv\bigl(\tau^{\mathrm{e\text{-}ph}}_{\nu\mathbf q}\bigr)^{-1}
= \frac{2\pi}{\hbar}\!
  \sum_{mn\mathbf k}
  \bigl|g^{\nu}_{mn}(\mathbf k,\mathbf q)\bigr|^{2}
  \bigl[f_{n\mathbf k}-f_{m,\mathbf{k+q}}\bigr]\,
  \delta\bigl(\epsilon_{n\mathbf k}-\epsilon_{m,\mathbf{k+q}}
             +\hbar\omega_{\mathbf q\nu}\bigr),
\label{eq:gamma_general}
\end{equation}
where the Pauli factor selects states near $E_{\mathrm F}$ and $g^{\nu}_{mn}$ are the DFPT/Wannier
matrix elements.\cite{Giustino2017_RMP,Sohier2014_PRB}  Before specializing the small $\omega$ scalings of the e-ph linewidths, it is useful to recall how the BTE integrand-the product of the phonon density of states \(D_\nu(\omega)\),
the modal heat capacity \(C_\nu(\omega)\), the squared group velocity \([v_\nu(\omega)]^{2}\),
and the lifetime \(\tau_\nu(\omega)\)—weights each branch in suspended graphene where the first-order ZA e-ph vertex is symmetry-forbidden. Full, iterative solutions of the phonon BTE (including normal processes) predict a \emph{ZA-dominated} lattice conductivity and a large gap relative to single-mode RTA. This baseline (ph–ph limited, $\sigma_h$ intact) is summarized in Fig. 5(a).

First, in-plane LA/TA modes. At long wavelength, graphene’s in–plane coupling is governed by an \emph{unscreened}
gauge–field (shear) interaction, while the scalar deformation potential is strongly screened.
First–principles BTE shows that this gauge term controls electronic resistivity and the
low–$q$ e–ph matrix elements for LA/TA modes.\cite{Sohier2014_PRB,Park2014_NanoLett}
Near charge neutrality the available electronic phase space is small, but for doped samples
$N(E_{\mathrm F})\!\propto\!|E_{\mathrm F}|$ grows linearly with $|E_{\mathrm F}|$ (or $k_F$) and
enhances $\Gamma_{\mathrm{LA/TA}}$ roughly in proportion to $k_F$ at fixed $\omega$.
Consequently, the LA/TA contribution to the e–ph–limited lattice conductivity
decreases with carrier density through the $k_F$–dependence of $\tau_{\mathrm{LA/TA}}$,
while retaining the weak $q$–dependence characteristic of linear modes.\cite{CastroNeto2009_RMP,DasSarma2011_RMP}

At long wavelength the \emph{gauge-field} (shear) coupling dominates and is only weakly $q$-dependent, whereas the scalar deformation potential is strongly screened.\cite{Sohier2014_PRB,Park2014_NanoLett}
In doped graphene this gives the small-$\omega$ scaling
\begin{equation}
\Gamma_{\mathrm{LA/TA}}(\omega)\;\propto\; k_F\,\omega
\quad\Rightarrow\quad
\tau_{\mathrm{LA/TA}}(\omega)\;\propto\;(k_F\,\omega)^{-1},
\label{eq:tau_LA_TA_scaling}
\end{equation}
consistent with the linear-$T$ acoustic resistivity of Dirac carriers.\cite{CastroNeto2009_RMP,DasSarma2011_RMP}

Second, flexural ZA modes and mirror symmetry.
For \emph{freestanding, mirror–symmetric} graphene the first–order (one–phonon)
e–ph coupling to ZA modes is symmetry–forbidden; ZA scattering of electrons then proceeds via
higher–order/two–phonon processes and is comparatively weak.\cite{Mariani2008_PRL}
This same mirror selection rule also restricts anharmonic ph–ph scattering of ZA modes,
which helps explain why ZA phonons dominate $\kappa_{\mathrm{ph}}$ in pristine
suspended graphene.\cite{Lindsay2014_PRB,Lindsay2010_PRB}
When mirror symmetry is \emph{broken} by a substrate, electrostatic gating, static ripples or
edges, a \emph{linear} one–phonon ZA coupling appears with small-$q$ amplitude $\propto q$, giving

\begin{equation}
\Gamma_{\mathrm{ZA}}(\omega)\;\propto\; k_F\,\omega^{1/2}
\quad\Rightarrow\quad
\tau_{\mathrm{ZA}}(\omega)\;\propto\;(k_F\,\omega^{1/2})^{-1},
\label{eq:tau_ZA_scaling}
\end{equation}

This greatly increases the ZA damping rate and reduces the ZA share of $\kappa_{\mathrm{ph}}$. First–principle studies show this mechanism can become mobility/thermal-transport limiting
in supported devices and under gating.\cite{Nika2017_RPP,Gunst2017_PRL}

In graphene the in-plane acoustic branches are linear,
$\omega_{\mathrm{LA/TA}}(q)=v_{s}\,q$, whereas the out-of-plane flexural branch is quadratic,
\begin{equation}
\omega_{\mathrm{ZA}}(q)=\sqrt{\kappa_b/\rho}\;q^{2},
\end{equation}
with bending rigidity $\kappa_b$ and areal mass density $\rho$.
In two dimensions the corresponding phonon DOS per unit area is
\begin{equation}
D_{\mathrm{LA/TA}}(\omega)=\frac{\omega}{2\pi v_s^{2}},\qquad
D_{\mathrm{ZA}}(\omega)=\frac{1}{4\pi}\sqrt{\frac{\rho}{\kappa_b}},
\end{equation}
with bending rigidity $\kappa_b$ and areal mass density $\rho$.
In two dimensions these dispersions imply that the phonon density of states
(DOS) scales as $D_{\mathrm{LA/TA}}(\omega)\!\propto\!\omega$ for linear modes
and is \emph{approximately constant} for the quadratic ZA branch
(contrary to a $\omega^{-1/2}$ scaling); this difference is central to the
low–frequency phase space in 2D.\cite{Gu2018_RMP,Nika2017_RPP,Lindsay2014_PRB}

Within the 2D BTE we have,

\begin{equation}
\kappa_{\mathrm{ph}}=\sum_{\nu}\int_{0}^{\omega_{\mathrm D}}\!\!d\omega\;D_{\nu}(\omega)\,C_{\nu}(\omega)\,v_{\nu}(\omega)^{2}\,\tau_{\nu}(\omega),
\end{equation}

and in the classical limit $C_{\nu}\!\approx\!k_B$.
The distinct DOS and group-velocity scalings mean that for LA/TA, $D_{\mathrm{LA/TA}}=\omega/(2\pi v_s^{2})$ and
$v\!\approx\!v_s$ yield an integrand that is roughly constant when $\Gamma_{\mathrm{LA/TA}}\!\propto\!\omega$.

\begin{equation}
\kappa_{\mathrm{LA/TA}}\;\propto\;\int_{\omega_{\min}}^{\omega_{\mathrm D}}
\omega\,[v_s^{2}]\,[k_F^{-1}\omega^{-1}]\,d\omega
\;=\;\frac{v_s^{2}}{k_F}\,\bigl(\omega_{\mathrm D}-\omega_{\min}\bigr),
\label{eq:kappa_LA_TA}
\end{equation}

so increasing $k_F$ primarily lowers
$\kappa_{\mathrm{LA/TA}}$ through shorter $\tau$. The infrared cutoff $\omega_{\min}$ set by finite size or tension.

For symmetry-broken ZA, $D_{\mathrm{ZA}}=\frac{1}{4\pi}\sqrt{\rho/\kappa_b}$ and $v_{\mathrm{ZA}}(\omega)=2\sqrt{\omega}\,(\kappa_b/\rho)^{1/4}$ so $v_{\mathrm{ZA}}^{2}\!\propto\!\omega$.
Using Eq.~(\ref{eq:tau_ZA_scaling}),
\begin{equation}
\kappa_{\mathrm{ZA}}\;\propto\;\int_{0}^{\omega_{\mathrm D}}
\!\Big[\tfrac{1}{4\pi}\sqrt{\tfrac{\rho}{\kappa_b}}\Big]\,[k_B]\,[\omega]\,[k_F^{-1}\omega^{-1/2}]\,d\omega
\;=\;\frac{\mathrm{const}}{k_F}\,\omega_{\mathrm D}^{3/2},
\label{eq:kappa_ZA}
\end{equation}
showing a stronger ultraviolet weighting than LA/TA. In pristine suspended graphene the linear ZA coupling is forbidden, $\Gamma_{\mathrm{ZA}}\!\to\!0$ at one-phonon level, and ZA modes dominate $\kappa_{\mathrm{ph}}$; on substrates or under gating, the activated ZA coupling enhances ZA damping and shifts heat carriage toward in-plane modes.\cite{Nika2017_RPP,Lindsay2014_PRB,Gunst2017_PRL} It should be noted that Eqs.~(\ref{eq:kappa_LA_TA})–(\ref{eq:kappa_ZA}) capture robust trends. Quantitative predictions require the full mode and $\mathbf q$ resolved $g^{\nu}_{mn}$, realistic low-$\omega$ cutoffs (finite size/tension for ZA), and the actual Debye (or numerical) upper limit.\cite{Giustino2017_RMP,Lindsay2014_PRB,Sohier2014_PRB}

\subsection{Effect of tensile strain on flexural modes and the thermal conductivity}

Applying a uniform uniaxial tensile strain $\epsilon$ couples the bending rigidity $\kappa_b$ to an in-plane membrane tension
$\sigma=Y\,\epsilon$ (with $Y$ the 2D Young’s modulus). The flexural (ZA) dispersion is renormalized from purely quadratic to\cite{Nika2017_RPP,Lindsay2014_PRB}

\begin{equation}
  \omega_{\mathrm{ZA}}^{2}(q,\epsilon)
  = \frac{\kappa_b}{\rho}\,q^{4} + \frac{\sigma}{\rho}\,q^{2},
\end{equation}

so that for \emph{small} wave vectors $q\ll q_{*}$, with

\begin{equation}
  q_{*}=\sqrt{\sigma/\kappa_b},
  \qquad
  v_{\epsilon}=\sqrt{\sigma/\rho},
\end{equation}

the branch becomes \emph{linear}, $\omega_{\mathrm{ZA}}\!\approx\! v_{\epsilon}\,q$; for
$q\gg q_{*}$ the bending term dominates and the quadratic law is recovered,
$\omega_{\mathrm{ZA}}\!\approx\!\sqrt{\kappa_b/\rho}\,q^{2}$. This has been shown in Fig. 5(b).

Two immediate consequences follows. First, low–$\omega$ DOS suppression. In 2D, the ZA phonon density of states (DOS) is \emph{constant} for the quadratic branch, but becomes $D(\omega)\!\propto\!\omega$ in the strain-linearised regime. Hence tensile strain depletes the low-frequency ZA DOS and reduces the phase space for low-$\omega$ ZA scattering.\cite{Gu2018_RMP,Nika2017_RPP}

Second, higher group velocity and modified scattering. Linearization raises the ZA group velocity from $v_{\mathrm{ZA}}=\partial\omega/\partial q\!\propto\!q$ (bending)
to the strain-controlled constant $v_{\epsilon}$ at small $q$.
Combined with the DOS change, this alters the three-phonon phase
space and lengthens ZA lifetimes in suspended graphene; within the BTE, larger $v^2\tau$ for ZA modes increases their contribution to $\kappa_{\mathrm{ph}}$.\cite{Lindsay2014_PRB}

First-principles BTE calculations show that even modest tensile strain $(\approx 1\%)$ linearizes ZA at long wavelengths and \emph{increases} the intrinsic lattice
thermal conductivity of suspended graphene; in the thermodynamic limit,
$\kappa_{\mathrm{ph}}$ becomes length-divergent under tension due to the altered
ZA scattering landscape.\cite{Lindsay2014_PRB}
For finite samples, the net gain depends on size and edge scattering: when the
characteristic length exceeds the dominant ZA mean-free paths (tens of microns and
beyond), strain yields sizable $\kappa_{\mathrm{ph}}$ enhancements; for smaller
membranes or ribbons, boundary scattering can mask the intrinsic increase, leading
to weaker strain dependence.\cite{Nika2017_RPP,Kuang2016_IJHMT}
Overall, strain-induced flexural hardening offers a practical lever to boost
thermal transport in large, suspended graphene, whereas in supported devices
mirror-symmetry breaking and substrate scattering reduce ZA lifetimes and can
counteract the strain benefit.\cite{Nika2017_RPP,Kuang2016_IJHMT}

\begin{figure}[h]
\centering
\includegraphics[width=0.95\columnwidth]{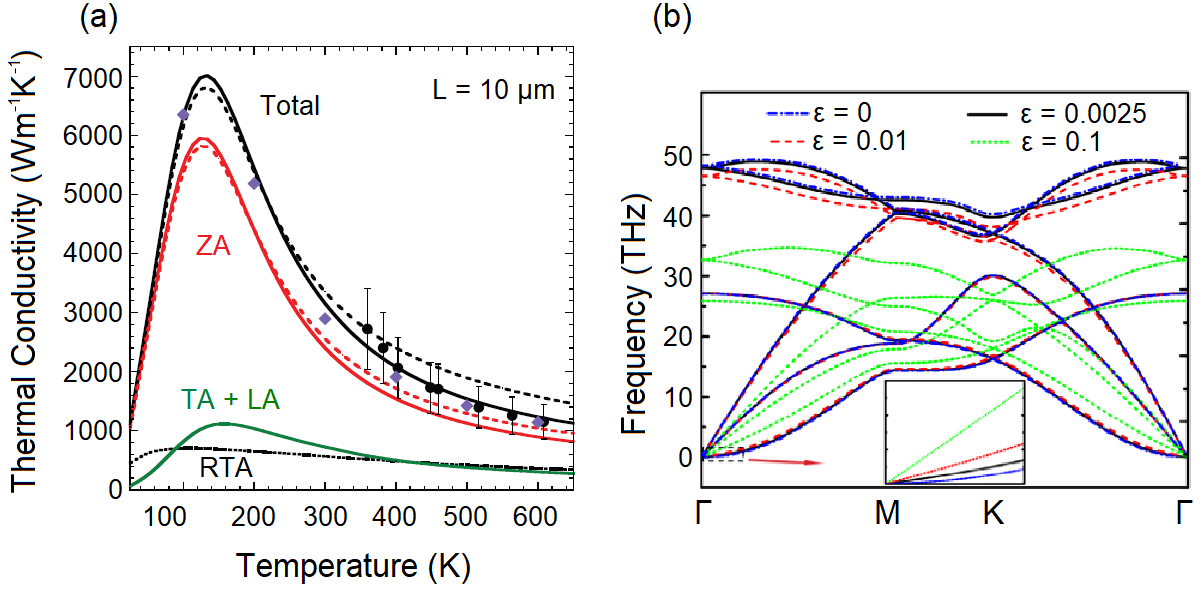}
\caption{(a) Branch-resolved lattice thermal conductivity of suspended monolayer graphene from first-principles BTE. The ZA branch dominates over the full T-range ($\approx 76 \%$ at 300 K). Experimental data over 9.7-µm holes has been shown for comparison. Figure taken from \emph{Lindsay et al.}\cite{Lindsay2014_PRB}. (b) Phonon dispersion under uniaxial tensile strain showing linearization of the ZA branch at small 
$q$ (and corresponding DOS change), which underpins the strain-enhanced $\kappa_{\mathrm ph}$ in suspended graphene. Figure taken from \emph{Kuang et al.}\cite{Kuang2016_IJHMT}.}
\label{Figure5}
\end{figure}

\subsection{Effect of first–order electron–phonon interactions on the lattice thermal conductivity}

At first-order, a single phonon scatters with a single electron, creating a new electron, EP $\rightarrow$ E$^{\ast}$. This adds an e-ph channel to the phonon collision operator, and in a phonon Boltzmann framework, this shortens the transport lifetime $\tau_{\nu\mathbf q}$ that controls $\kappa_{\mathrm{ph}}$.\cite{yang2021indirect} Using the standard decoupled, linearised phonon BTE, the lattice thermal conductivity can be written as
\begin{equation}
\kappa_{\mathrm{ph}}
=\frac{1}{N_{\mathbf q}\,V\,k_B T^{2}}
\sum_{\nu\mathbf q}
n^0_{\nu\mathbf q}\!\left(n^0_{\nu\mathbf q}{+}1\right)\,
(\hbar\omega_{\mathbf q\nu})^{2}\,
\mathbf v_{\nu\mathbf q}\!\cdot\!\mathbf F_{\nu\mathbf q},
\label{eq:kappa_master}
\end{equation}
where $n^0_{\nu\mathbf q}$ is the Bose factor, $\omega_{\mathbf q\nu}$ and $\mathbf v_{\nu\mathbf q}$ are the phonon frequency and group velocity, and $\mathbf F_{\nu\mathbf q}$ is the BTE response vector (in the relaxation–time approximation, $\mathbf F_{\nu\mathbf q}=\tau_{\nu\mathbf q}\,\mathbf v_{\nu\mathbf q}$).\cite{yang2021indirect} According to Matthiessen’s rule, the total mode lifetime is commonly assembled as

\begin{equation}
(\tau_{\nu\mathbf q})^{-1}
=(\tau_{\nu\mathbf q}^{\mathrm{ph-ph}})^{-1}
+(\tau_{\nu\mathbf q}^{\mathrm{e-ph}})^{-1}
+(\tau_{\nu\mathbf q}^{\mathrm{iso}})^{-1}
+(\tau_{\nu\mathbf q}^{\mathrm{B}})^{-1},
\end{equation}

where $\tau_{\nu\mathbf q}^{\mathrm{iso}}$ is the isotropic scattering lifetime and $\tau_{\nu\mathbf q}^{\mathrm{B}}$ is the boundary scattering lifetime. We derive the \emph{first–order} e–ph rate, Eq.~\eqref{eq:gamma_general}, using the DFPT/Wannier e–ph matrix elements $g^{\nu}_{mn}(\mathbf k,\mathbf q)$.\cite{Giustino2017_RMP,Sohier2014_PRB,Yue2019_PRB_SiPh}

In suspended, mirror–symmetric graphene, the one–phonon coupling to flexural ZA modes is symmetry–forbidden at first-order, so the direct e-ph channel acts primarily on the in–plane LA/TA branches.\cite{Kaasbjerg2012_PRB_GrapheneAcoustic} Iterative (decoupled) phonon–BTE calculations nevertheless find a \emph{large net reduction} of $\kappa_{\mathrm{ph}}$ when the first–order $(\tau_{\nu\mathbf q}^{\mathrm{e-ph}})^{-1}$ is included, because frequent LA/TA–electron events feed back—via abundant normal phonon processes—into the \emph{transport} lifetimes of all branches (including ZA).\cite{yang2021indirect} Quantitatively, turning on the first-order e-ph channel yields a $\sim\!21\%$ drop of $\kappa_{\mathrm{ph}}$ at 300 K and $\sim\!32\%$ at $200$~K, with a non–monotonic carrier–density trend exhibiting a minimum near $n\simeq 4.9\times10^{14}\,\mathrm{cm^{-2}}$.\cite{yang2021indirect} This can be seen in Fig. 6(a,c). Furthermore, for graphene, the long–wavelength LA/TA e–ph matrix elements admit analytic forms that match first–principles results and govern the temperature/doping dependence of acoustic e–ph scattering across the Bloch–Gr\"uneisen crossover; these forms feed directly into the $\mathbf q$– and $\omega$–dependence of $(\tau_{\nu\mathbf q}^{\mathrm{e-ph}})^{-1}$ used in Eq.~\eqref{eq:gamma_general} \cite{Kaasbjerg2012_PRB_GrapheneAcoustic} and is shown in Fig. 6(b,d).

First–principles DFPT/EPW workflows applied to other 2D Dirac crystals, such as silicene, explicitly compute $(\tau_{\nu\mathbf q}^{\mathrm{e-ph}})^{-1}$ for each branch (ZA/TA/LA) as functions of frequency and carrier density. Adding these mode–resolved e-ph rates to the intrinsic ph–ph terms (Matthiessen assembly) and solving the phonon BTE iteratively demonstrates a sizable \emph{suppression} of $\kappa_{\mathrm{ph}}$ under doping: the first–order e-ph channel trims the long mean–free–path tail and reduces branch–resolved conductivities in a manner consistent with symmetry and phase–space constraints.\cite{Yue2019_PRB_SiPh}

The three takeaways of first-order e-ph interactions in 2D Dirac crystals are:
(i) \emph{Mechanism}: the first-order e–ph channel adds $(\tau_{\nu\mathbf q}^{\mathrm{e-ph}})^{-1}$ that shortens transport lifetimes in the BTE integrand; because normal phonon processes are strong in graphene, LA/TA–electron scattering indirectly reduces ZA \emph{transport} lifetimes as well.\cite{yang2021indirect}
(ii) \emph{Magnitude}: in graphene this yields $\sim$20–30\% reductions of $\kappa_{\mathrm{ph}}$ in the 200–300 K range at finite carrier densities.\cite{yang2021indirect}
(iii) \emph{Generality}: comparable, explicitly first–order reductions are found in other 2D crystals (silicene, phosphorene) once $(\tau_{\nu\mathbf q}^{\mathrm{e-ph}})^{-1}$ is included and the BTE is solved beyond RTA.\cite{Yue2019_PRB_SiPh}

\begin{figure}[h]
\centering
\includegraphics[width=0.95\columnwidth]{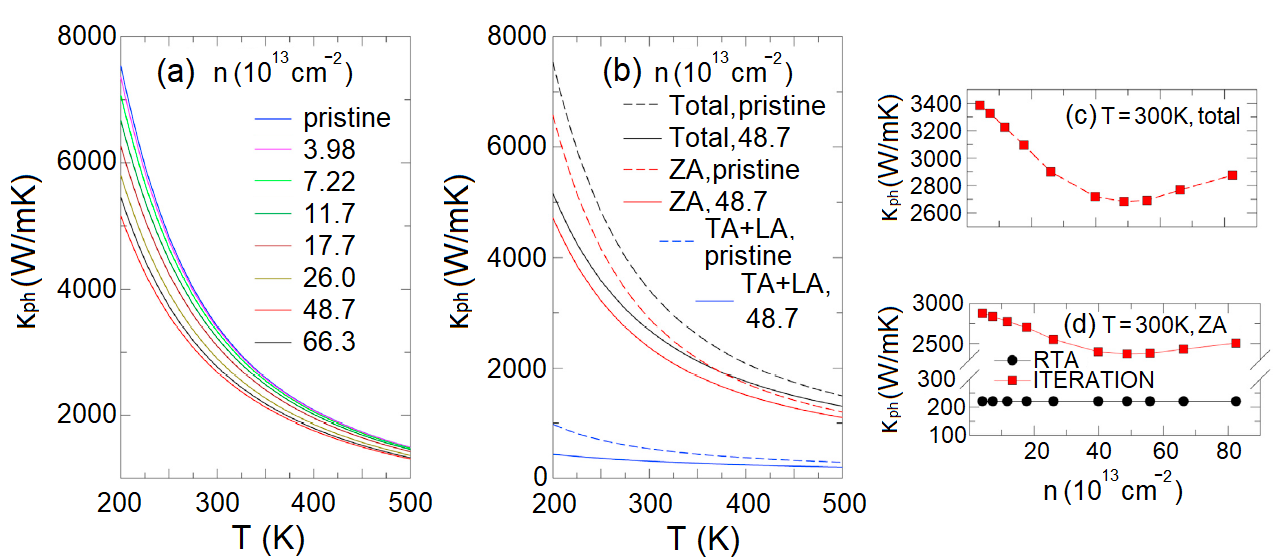}
\caption{Graphene lattice thermal conductivity with first-order e-ph scattering. Panels (a,c): total lattice thermal conductivity $\kappa_{\mathrm{ph}}$ from an iterative, decoupled phonon-BTE including the first-order e-ph channel. 
(a) $\kappa_{\mathrm{ph}}(T)$ for several carrier densities $n$ shows a larger e-ph impact at lower $T$; 
(c) $\kappa_{\mathrm{ph}}(n)$ at fixed $T$ exhibits a non-monotonic trend with a minimum near $n\!\approx\!4.9\times10^{14}\,\mathrm{cm^{-2}}$. 
Panels (b,d): mechanism and transport treatment. 
(b) Branch-resolved $\kappa_{\mathrm{ph}}$ (LA/TA vs. ZA) demonstrates that while the one-phonon ZA coupling is symmetry-forbidden in suspended graphene, frequent LA/TA electron events (together with abundant normal phonon processes) indirectly suppress the ZA transport contribution; 
(d) Comparison of relaxation-time approximation (RTA) with the fully iterative solution highlights RTA’s underestimate and the importance of normal-process (vertex) corrections. 
Together, the panels quantify how first-order e-ph scattering shortens transport lifetimes, trims the long-MFP tail, and redistributes heat among branches even when direct ZA–electron coupling vanishes by symmetry. Figure taken from \emph{Yang et al.}\cite{yang2021indirect}.}
\label{Figure6}
\end{figure}

\subsection{Screening, carrier density, and flexural vs.\ in-plane scattering in graphene.}

In lightly doped or gated graphene ($n\!\lesssim\!10^{12}\,\mathrm{cm^{-2}}$), the static screening wave vector grows as $q_s\!\propto\!k_F\!\propto\!\sqrt{n}$ and is therefore small.\cite{CastroNeto2009_RMP,DasSarma2011_RMP}
In this regime the \emph{gauge-field} (shear) coupling that governs long-wavelength LA/TA scattering is essentially \emph{unscreened}, while scalar (deformation-potential–like) terms are only weakly screened.\cite{Sohier2014_PRB,Park2014_NanoLett}
If mirror symmetry is broken (by a substrate, gating, ripples, or edges), a \emph{linear} one-phonon coupling to flexural (ZA) modes is activated with small-$q$ amplitude $\propto q$;\cite{Mariani2008_PRL,Gunst2017_PRL}
under those conditions the ZA-induced electronic scattering can become comparable to the LA/TA channel at room temperature for low $n$.
First-principles electron BTE calculations (beyond RTA) find that allowing symmetry-broken, one-phonon ZA processes reduces the room-temperature mobility by on the order of tens of percent in otherwise clean devices, with a weak $n$-dependence for $n\!\lesssim\!10^{12}\,\mathrm{cm^{-2}}$.\cite{Sohier2014_PRB,Gunst2017_PRL}
The weak density trend reflects two facts: (i) at small $k_F$ the static screening of scalar terms is modest, and (ii) the dominant low-$q$ matrix elements scale only linearly with $q$, so the overall phase-space weighting does not change dramatically with $n$ in this limit.\cite{DasSarma2011_RMP,Sohier2014_PRB}

At higher doping ($n\!\gtrsim\!10^{13}\,\mathrm{cm^{-2}}$), $q_s$ exceeds typical thermal phonon wave vectors ($q\!\sim\!k_BT/\hbar v_s$ for LA/TA) and strongly suppresses scalar (deformation-potential–like) contributions—including the symmetry-allowed ZA coupling—except in the narrow backscattering window near $q\!\approx\!2k_F$.\cite{CastroNeto2009_RMP,DasSarma2011_RMP}
The in-plane acoustic channel, dominated by the unscreened gauge-field coupling, therefore regains clear dominance.\cite{Sohier2014_PRB,Park2014_NanoLett}
In this degenerate regime the intrinsic acoustic-phonon-limited resistivity remains approximately linear in $T$ and only weakly dependent on $n$; consequently, the \emph{conductivity} $\sigma$ grows roughly $\propto n$ (at fixed $T$), while the \emph{mobility} $\mu=\sigma/(ne)$ is nearly $n$-independent.\cite{DasSarma2011_RMP,Park2014_NanoLett}
By Wiedemann–Franz, the electronic thermal conductivity then scales as $\kappa_{\mathrm e}\!\approx\!L_0 T\,\sigma$ and increases with $n$ through $\sigma$, whereas flexural-phonon contributions to electronic scattering become a negligible correction in clean, highly doped samples.\cite{DasSarma2011_RMP}

\subsection{Size, edge roughness, and specularity}

In strictly $\sigma_h$-symmetric monolayers the one-phonon ZA/ZO e–ph matrix elements vanish; a \emph{linear} ZA coupling is activated only when $\sigma_h$ is broken (e.g., by substrates, gating, ripples, or edges). In micron-scale graphene ribbons, intrinsic (ph–ph) mean-free paths (MFPs) are strongly branch dependent. First-principles BTE studies show that, in suspended graphene at 300 K, the in-plane LA/TA contributions saturate for MFPs of order $\sim\!10~\mu$m or less, whereas the flexural ZA contribution accumulates out to \emph{centimeter}-scale MFPs (ultra-long tails in the MFP spectrum).\cite{Lindsay2014_PRB,Kuang2016_IJHMT} This has been displayed in Fig. 7(a). When the ribbon width $W$ becomes comparable to, or smaller than, these intrinsic lengths, boundary scattering must be included. A standard Ziman/Casimir specularity correction gives an edge-scattering rate

\begin{equation}
  \tau_{\nu}^{-1}\bigl(\text{edge}\bigr)
  \;=\;\frac{v_{\nu}}{\Lambda_B}
  \;=\;\frac{v_{\nu}}{W}\,\frac{1-P}{1+P},
\end{equation}

with $P$ the specularity parameter ($P\!=\!1$ mirror-like, $P\!=\!0$ fully diffuse) and $v_{\nu}$ the branch group velocity.\cite{Shen2014_JAP}
This has been shown in Fig. 7(b). Because $v_{\mathrm{LA}}$ and $v_{\mathrm{TA}}$ exceed the small-$q$ flexural velocity $v_{\mathrm{ZA}}(q)=\partial\omega_{\mathrm{ZA}}/\partial q\propto q$ by an order of magnitude, the \emph{same} edge roughness $(1-P)$ imposes a larger additional rate on the in-plane modes.

\begin{figure}[h]
\centering
\includegraphics[width=0.95\columnwidth]{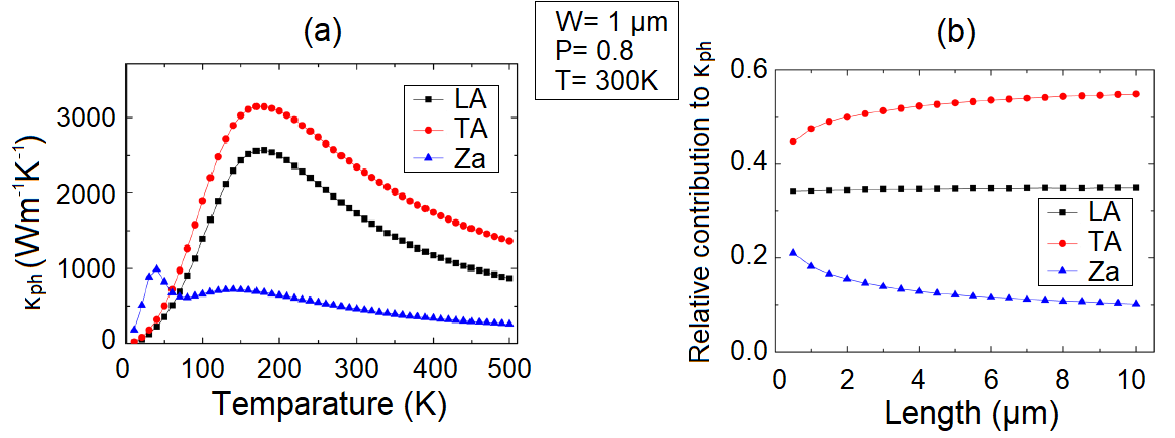}
\caption{Panel (a): Branch-resolved cumulative lattice thermal conductivity vs mean-free-path in suspended graphene at 300 K.
LA/TA contributions saturate for \(\Lambda \lesssim 10~\mu\)m, whereas the flexural ZA branch exhibits an ultra-long tail extending to \(\gtrsim\)cm scales. This hierarchy explains why, when the ribbon width \(W\) approaches a few microns, boundary/specularity scattering predominantly suppresses in-plane modes, while ZA remains comparatively resilient—consistent with the discussion in the main text. Panel (b): Relative LA/TA/ZA contributions to $\kappa_{\mathrm ph}$ vs ribbon length L (fixed width/specularity). The boundary scattering suppresses in-plane modes first, increasing the ZA share as L shortens. Figures taken from \emph{Shen et al.}\cite{Shen2014_JAP}.}
\label{Figure7}
\end{figure}

Quantitatively, linearised BTE simulations for a $W=1~\mu$m ribbon show that reducing $P$ from $0.9$ to $0.3$ lowers the LA contribution to $\kappa_{\mathrm{ph}}$ by $\approx 40\%$ and TA by $\approx 45\%$, while the ZA contribution drops by only $\approx 20\%$; hence the flexural share of the lattice heat flow \emph{increases} as edges become rougher.\cite{Shen2014_JAP}
Length scaling reinforces this picture: for $L<1~\mu$m the ZA branch already contributes $>\!20\%$ of $\kappa_{\mathrm{ph}}$ (even with moderately specular edges), and as $L$ is reduced further into the few-hundred-nanometer regime, boundary scattering overwhelms the in-plane branches first, leaving ZA modes as the principal heat carriers unless mirror symmetry is broken by a substrate.\cite{Shen2014_JAP}
In suspended, unstrained graphene this outcome is amplified by symmetry-based selection rules that extend ZA lifetimes; coupling to a substrate or applying tension suppresses ZA dominance by breaking mirror symmetry and/or modifying the ZA dispersion.\cite{Lindsay2014_PRB,Kuang2016_IJHMT}

Overall, rough or narrow geometries preferentially attenuate the fast in-plane LA/TA carriers far more than the slower but densely populated ZA modes, making the ZA branch an efficient thermal channel in graphene nanoribbons over a wide range of sizes and edge specularities.\cite{Lindsay2014_PRB,Shen2014_JAP,Kuang2016_IJHMT}

\subsection{The role of higher order electron-phonon interactions}

In 2D Dirac crystals (e.g., graphene) the band dispersion is linear, so the Fermi energy scales as \(E_{\mathrm F}=\hbar v_F k_F\).
Because viable gate–induced sheet densities are modest, \(k_F\) is small and \(E_{\mathrm F}\) typically lies in the \(50\text{--}200~\mathrm{meV}\) range, far below the multi–eV Fermi energies of ordinary metals.
As a result, at elevated temperatures \(k_B T\sim 0.05\text{--}0.1~\mathrm{eV}\) (600–1200 K) we have \(E_{\mathrm F}\!\not\!\gg\!k_B T\).
In this regime two Dirac–specific features become important.
First, screening cannot be treated with Thomas–Fermi: the \emph{dynamic} Lindhard response to \emph{acoustic} phonons exhibits a sharp cusp at \(q=2k_F\) even in the static limit and further modifies the coupling in the dynamical regime, so phonons with momenta near \(2k_F\) are strongly affected~\cite{kazemian2023dynamic}.
This is a behavior missed by Thomas–Fermi and characteristic of low–\(E_{\mathrm F}\) Dirac bands.
Second, because \(E_{\mathrm F}\) is low, the three–particle process composed of the \emph{annihilation of an electron–phonon pair and the creation of a new electron}, \(\mathrm{EP}\!\to\!\mathrm{E}^{\ast}\), no longer overwhelmingly dominates over the three–particle process in which \emph{an electron decays into an electron–phonon pair}, \(\mathrm{E}\!\to\!\mathrm{E}^{\ast}\mathrm{P}^{\ast}\).
Their rates become comparable at low Fermi energies and high temperatures and can partially cancel.
This cancellation reduces the \emph{net} first–order e–ph damping at high \(T\) and low \(E_{\mathrm F}\), necessitating inclusion of higher–order (four–particle) processes such as \(\mathrm{EP}\!\leftrightarrow\!\mathrm{E}^{\ast}\mathrm{P}^{\ast}\), which in certain regimes provide the leading contribution to the e–ph–limited phonon lifetime and to the electron–related part of the lattice thermal conductivity~\cite{taylor2002quantum,kazemian2024influence}.
(The four–particle process with the presence of a fermion propagator \(P_F\) is shown in Fig. 8(a).)

In the four–particle channel \(\mathrm{EP}\!\leftrightarrow\!\mathrm{E}^{\ast}\mathrm{P}^{\ast}\), crystal momentum is conserved \emph{modulo} a reciprocal–lattice vector \(\mathbf G\) (Umklapp).
Whenever the created phonon lies outside the first Brillouin zone its wave vector folds back as \(\mathbf q''-\mathbf G\).
A simple enumeration of neighboring zones shows that roughly one half of the nominally allowed \(\mathbf q''\) fall into Umklapp–active sectors where \(|\mathbf q''-\mathbf G|\!\ll\!|\mathbf q''|\); this is depicted in Fig. 8(b).
These events divert electronic momentum into lattice recoil and effectively \emph{remove} about half of the kinematically available outgoing–phonon phase space for \(\mathrm{EP}\!\leftrightarrow\!\mathrm{E}^{\ast}\mathrm{P}^{\ast}\), thereby lowering the corresponding four–particle rate and imposing an intrinsic cap on the associated phonon mean–free path~\cite{kazemian2024influence}.
Taken together, (i) the need for a Lindhard (not TF) dielectric with a \(2k_F\) cusp for acoustic phonons, and (ii) the rising importance of \(\mathrm{E}\!\to\!\mathrm{E}^{\ast}\mathrm{P}^{\ast}\) and \(\mathrm{EP}\!\leftrightarrow\!\mathrm{E}^{\ast}\mathrm{P}^{\ast}\) when \(E_{\mathrm F}\!\sim\!k_B T\), are derived and quantified specifically for 2D Dirac crystals.

\textbf{Scope and working assumption.} In what follows we \emph{restrict} the analysis to (i) strictly monolayer Dirac crystals that possess exact horizontal–mirror symmetry \(\sigma_h\) with all atoms in the mirror plane, and (ii) graphene nanoribbons (GNRs) in device–relevant sizes. In case (i) the deformation potential generated by a purely out–of–plane displacement is odd under \(\sigma_h\), so the first–order \emph{intraband} e–ph matrix elements for flexural acoustic/optical modes (ZA/ZO) vanish identically,
\[
\bigl\langle\psi_{n\mathbf k}\big|\,\partial \phi_{\mathrm{ZA/ZO}}\,\big|\psi_{n\mathbf k}\bigr\rangle = 0,
\]
and the e–ph–limited transport is governed by the \emph{in–plane} acoustic branches (LA/TA), consistent with the classic in–plane graphite treatment~\cite{Mariani2008_PRL,Alidoosti2022_sigmah,Klemens2000_GraphiteAplane}. In case (ii) (GNRs), size and edge roughness renormalize branch contributions: diffusive edges and finite length suppress LA/TA more strongly than ZA, so the \emph{relative} ZA share of \(\kappa_{\mathrm{ph}}\) can increase as ribbons narrow or edges roughen (see Fig. 6(a)); nonetheless, for long, high–quality ribbons with nearly specular edges the LA/TA modes remain the dominant heat carriers~\cite{Shen2014_JAP,Mariani2008_PRL,Nika2017_RPP,Lindsay2014_PRB,Kuang2016_IJHMT,liu2014anomalous}. Accordingly, the four–particle channel \(\mathrm{EP}\!\leftrightarrow\!\mathrm{E}^{\ast}\mathrm{P}^{\ast}\) that we formulate below is written explicitly for the LA/TA branches. Flexural (ZA/ZO) contributions are excluded by symmetry in \(\sigma_h\)–preserving monolayers, and are subdominant under the ribbon conditions we emphasize here.

\begin{figure}[h]
\centering
\includegraphics[width=.9\columnwidth]{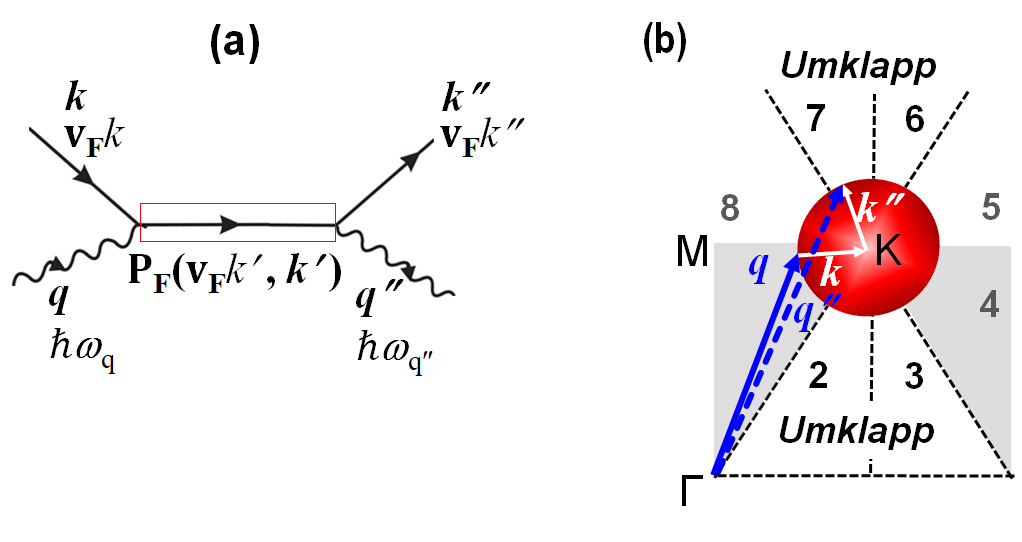}
\caption{(a) The four-particle e-ph interaction consisting of the annihilation of a pair of electrons and phonons and the creation of a new pair, \(\mathrm{EP}\!\leftrightarrow\!\mathrm{E}^{\ast}\mathrm{P}^{\ast}\), with a Fermion propagator in the middle. (b) The wave vector of the outgoing phonon scattered to the 1st, 4th, 5th, and 8th neighboring lattice sites remain approximately the same while the wave vector of the phonon scattered to the 2nd, 3rd, 6th, and 7th neighboring lattice sites will be much smaller than the annihilated phonon since its wave vector will be hugely affected by the Umklapp process rendering the process inelastic. Figure taken from \emph{Kazemian et al.}\cite{kazemian2024influence}.}
\label{Figure8}
\end{figure}

The interaction Hamiltonian of the four-particle e-ph process shown is written as:

\begin{equation}
\begin{split}
   H^{4}_{e-ph}=\sum_{q,q''} \sum_{k,k',k''} &\sqrt{\frac{\hbar^{2}}{2m\hbar \omega_{\textbf{q}}}}\cdot q \; \phi_s(q) \; b_{q} c^{\dagger}_{k'} c_{k} \; e^{i(\textbf{k}'-\textbf{k}- \textbf{q})\cdot \textbf{r}} \; e^{-\frac{i}{\hbar}(E_{k}'-E_{k}- \hbar \omega_{\textbf{q}})\cdot t} P_{F}(E_{k'},k') \\& \sqrt{\frac{\hbar^{2}}{2m\hbar \omega_{\textbf{q}''}}}\cdot q'' \; \phi_s(q'') \;b^{\dagger}_{q''} c^{\dagger}_{k''} c_{k'} \; e^{-i(\textbf{k}''-\textbf{k}'- \textbf{q}'')\cdot \textbf{r}} \; e^{\frac{i}{\hbar}(E_{k''}-E_{k}- \hbar \omega_{\textbf{q}''})\cdot t},
\label{The interaction Hamiltonian of the four-particle e-ph process}
\end{split}
\end{equation}

where $P_F(E_{k'},k')$ is the Fermion propagator. We can then write the phonon scattering rate for the 4-particle process as 

\begin{equation}
\begin{split}
    \frac{1}{\tau^{EP-E^{*}P^{*}}}= &  \frac{1}{2} \Bigg( \frac{16\pi^4 \alpha^{4}}{\hbar} \Bigg) \sum^{q_{max}}_{q_{min}} \;
    \sum^{q''_{max}}_{q''_{min}} \sum^{k_{\mathrm{F}}}_{k,k''=0}   \; \Big<f_{k}(1-f_{k''})\Big> \Big<n_q\Big> \; \delta(E_{k}+cq-E_{k''}-cq''\big) \\& \Big(\frac{1}{q-2\pi\chi(q)}\Big)^{2} \frac{q \; q''}{\Bigg[\Big(k^{2}+q^{2}-2kq\cos{\theta}\Big)^{1/2}-k-(c/v_{\mathrm{F}})q\Bigg]^{2}}\Big(\frac{1}{q''-2\pi\chi(q'')}\Big)^{2}.
\label{phonon scattering rate of four-particle process}
\end{split}
\end{equation} 

where is the speed of sound and $\theta$ is the angle of collision between the annihilated phonon and electron and $\chi(q)$ is the dynamic Lindhard dielectric response function.\cite{kazemian2023dynamic} 
Knowing the e-ph lifetime $\tau_{(e-ph)}$ one can easily calculate the e-ph thermal conductivity

\begin{equation}
\begin{split}
    k_{th,(e-ph)}=\frac{1}{2}v^{2}_{gr} \sum_{q} C_{q}\tau_{(e-ph)},
\label{e-ph thermal conductivity}
\end{split}
\end{equation}

where $v_{gr}$ is the group velocity of acoustic phonons in the 2D Dirac crystal, and $C_{q}$ is the specific heat capacity per unit area. The e-ph thermal conductivity for in-plane phonons is depicted in Fig. 9 at two different temperatures. Incorporating the four-particle process in the analysis of e-ph thermal conductivity results in a notably slower increase in slope at small Fermi energies compared to considering only the three-particle process, emphasizing the significance of higher-order e-ph interactions in studying 2D Dirac crystals, particularly when examining low Fermi energies.

\newpage

\begin{figure}[h]
\centering
\includegraphics[width=.9\columnwidth]{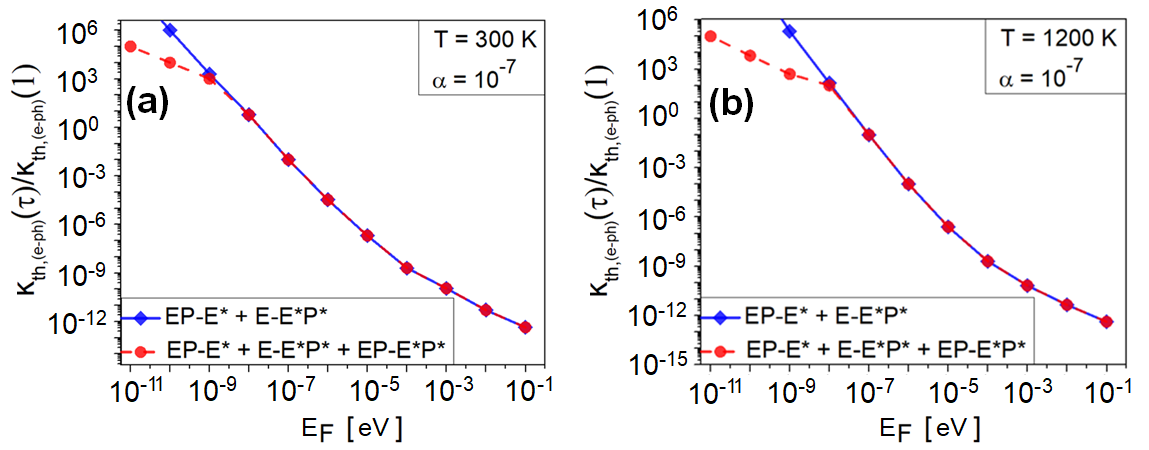}
\caption{(a,b) The e-ph thermal conductivity with respect to the Fermi energy of a 2D Dirac crystal with $\alpha=10^{-7}$. The variable $\alpha$ is dependent on the unit cell area, mass, and the ion charge of the 2D Dirac crystal. For a 2D Dirac crystal such as graphene with a unit cell area of $\Omega \approx 3.2*10^{-16}$ we have $\alpha \approx 10^{-7}$. We observe that when we incorporate the four-particle process in the calculation of the e-ph thermal conductivity, the rate of increase in slope is significantly reduced at lower Fermi energies compared to when only the three-particle process is considered. This finding emphasizes the significance of higher-order e-ph interactions in the investigation of 2D Dirac crystals, particularly at low Fermi energies. Figure taken from \emph{Kazemian et al.}\cite{kazemian2024influence}.}
\label{Figure9}
\end{figure}
\raggedbottom

\subsection{Twisted graphene}

Twist-engineered graphene hosts flat moiré minibands with enhanced density of states and modified screening, which can amplify e-ph interactions and open new scattering channels set by moiré reciprocal vectors. Microscopically, several works have shown that phonons can mediate appreciable intervalley attraction in magic-angle twisted bilayer graphene (MATBG), potentially supporting superconductivity with $s$- or $d$-wave character depending on competing interactions\cite{Wu2018_PRL_TBGSC,Lian2019_PRL_TBGSC}. Atomistic calculations further indicate strong e-ph coupling, pronounced electron-hole asymmetry, and nonadiabatic effects in MATBG\cite{Choi2018_PRB_TBG_EPC}. Recent experiments report direct signatures of strong intervalley e-ph coupling in MATBG, consistent with these theoretical expectations\cite{Chen2024_Nature_StrongEPC}.

On the lattice side, the moiré superlattice reconstructs acoustic/optical branches into \emph{moiré phonon} bands with mini-gaps and collective domain-wall vibrations, and similar effects arise in aligned graphene/hexagonal boron nitride (h-BN) heterostructures\cite{Koshino2019_PRB_MoirePhonons,Krisna2023_PRB_GhBN_MoirePhonons}. These modes reshuffle the available phase space for e-ph and ph-ph scattering, enable “moiré-Umklapp” momentum relaxation via small wave vectors, and can couple efficiently to flat-band carriers. In supported devices, \emph{remote interfacial phonons} from substrates (e.g., h-BN/oxides) may further influence resistivity and hot-phonon dynamics; accounting for these environment-dependent channels is increasingly important in quantitative transport.

From a thermal-transport standpoint, moiré systems sit at the interface between simple crystals and complex unit cells with many near-degenerate modes. In regimes where phonon linewidths become comparable to interbranch splittings, coherences/tunneling contributions to $\kappa$ (beyond Peierls populations) may become relevant. The unified formulation $\kappa=\kappa^{\mathrm P}+\kappa^{\mathrm C}$ provides the appropriate extension in such cases (see our Sec.~1.2.2 note)~\cite{Simoncelli2019_NatPhys}.

\subsection{Summary and open challenges}

Carrier density, lattice symmetry, strain, and finite geometry jointly dictate how e-ph interactions limit transport in 2D Dirac crystals. At low carrier densities ($n\!\lesssim\!10^{12}\,\mathrm{cm}^{-2}$), screening is weak ($q_s\!\propto\!k_F$), linear one-phonon flexural (ZA) coupling is activated whenever mirror symmetry is broken (by substrates, curvature, or edges), and ZA-induced electronic scattering can become comparable to the in-plane LA/TA channel, reducing intrinsic mobility by on the order of tens of percent in otherwise clean devices.\cite{Mariani2008_PRL,Sohier2014_PRB,Gunst2017_PRL} At higher densities ($n\!\gtrsim\!10^{13}\,\mathrm{cm}^{-2}$), Thomas-Fermi/Lindhard screening suppresses scalar (deformation-potential-like) terms and restricts backscattering to a narrow window near $q\!\approx\!2k_F$, restoring in--plane dominance. In this degenerate regime  In degenerate, acoustic-phonon-limited graphene the resistivity is approximately linear in $T$ and only weakly dependent on $n$; hence at fixed $T$ the conductivity is roughly $n$-independent (and the mobility scales as $\mu\!\propto\!1/n$), unless additional mechanisms (e.g.\ screening of a scalar deformation potential or multi-channel scattering) introduce an $n$-dependence.\cite{DasSarma2011_RMP,Park2014_NanoLett,Sohier2014_PRB} 
Uniform tensile strain introduces an in-plane tension $\sigma=Y\epsilon$ that linearizes the ZA dispersion below a crossover $q_\ast=\sqrt{\sigma/\kappa_b}$, converting the constant (quadratic) ZA density of states into $D(\omega)\!\propto\!\omega$, raising the long--wavelength ZA group velocity, lengthening ZA lifetimes, and increasing the intrinsic lattice thermal conductivity of suspended membranes (with the gain ultimately limited by sample size and boundaries).\cite{Nika2017_RPP,Lindsay2014_PRB,Kuang2016_IJHMT} If horizontal--mirror symmetry ($\sigma_h$) is exact and all atoms lie in the mirror plane (e.g. free-standing graphene), the first-order ZA/ZO e-ph matrix elements vanish. Once symmetry is broken by edges or substrates, linear ZA coupling reappears. In graphene nanoribbons, boundary scattering modeled via Ziman/Casimir specularity suppresses LA/TA strongly and can raise the \emph{relative} ZA share as ribbons shorten or edges roughen, although in long, high-quality ribbons the in-plane branches typically remain the dominant heat carriers.\cite{Nika2017_RPP,Shen2014_JAP,Alidoosti2022_sigmah} Furthermore, it has been shown that in 2D Dirac crystals, at elevated temperatures and low Fermi energies, the E $\rightarrow$ E$^{\ast}$P$^{\ast}$ process has the potential to partially or entirely nullify the EP $\rightarrow$ E$^{\ast}$ process. It therefore becomes imperative to investigate higher order e-ph interactions in order to comprehend the overall dynamics accurately. Findings demonstrate that the impact of the four-particle process, EP$\leftrightarrow$ E$^{\ast}$P$^{\ast}$, on both phonon scattering rate and e-ph thermal conductivity. Umklapp further prunes the available phase space in this four-particle channel, modifying phonon lifetimes and the e-ph contribution to $\kappa$.\cite{kazemian2024influence}

Looking forward, several challenges remain \emph{within Dirac crystals}. A first priority is to extend coupled e-ph transport formalisms to time-dependent drives in the ultrafast regime of hot carriers and nonthermal phonons, while enforcing Kelvin-Onsager reciprocity. For graphene this requires marrying first-principles e-ph matrix elements with nonequilibrium carrier dynamics and the symmetry constraints specific to ZA/ZO modes.\cite{Giustino2017_RMP,CastroNeto2009_RMP,Alidoosti2022_sigmah} Second, beyond-RTA, fully iterative/variational solutions of the Peierls-Boltzmann equation with mode-to-mode couplings and higher order scattering are needed to quantify drag and boundary effects as device dimensions approach intrinsic mean--free paths.\cite{Lindsay2014_PRB} Third, accurate e-ph workflows for correlated or magnetic \emph{Dirac} crystals will require beyond-DFT electronic structures (hybrid functionals, $GW$, or DMFT) so that EPC rests on realistic quasiparticles.\cite{Giustino2017_RMP,CastroNeto2009_RMP} Fourth, systematic experimental benchmarks on \emph{graphene and related Dirac membranes} TDTR and micro-Raman on ribbons and strained, suspended stacks with \emph{known} edge specularity, curvature, substrate dielectric environment, and carrier density are still scarce and essential to validate higher order and coupled BTE predictions.\cite{Nika2017_RPP} Fifth, device scale multiphysics that embeds first principles e-ph lifetimes (including ZA symmetry rules and higher-order channels) into electro-thermal solvers is needed to translate these microscopic insights into quantitative design rules for Dirac transistors, interconnects, and thermoelectric elements.\cite{kazemian2024influence,Nika2017_RPP} Finally, for twisted (moiré) graphene, a practical route to quantitative charge and heat transport is to combine \emph{DFPT} with Wannier–interpolation (e.g., EPW) to obtain e–ph matrix elements on dense meshes in the moiré Brillouin zone, together with \emph{moiré phonon} calculations that capture band folding, mini–Umklapp processes, and possible remote interfacial phonons from encapsulation/substrates. Transport can then be treated within the Peierls–BTE (population) framework and, where quasi-degenerate branches make coherences relevant, augmented by the unified/coherence correction $\kappa=\kappa^{\mathrm P}+\kappa^{\mathrm C}$. A fully rigorous microscopic picture in these low-dimensional Dirac systems further benefits from (i) environment-dependent screening (gates/encapsulation), (ii) nonadiabatic effects associated with flat bands, (iii) higher-order anharmonicity (four-phonon) at elevated temperatures, and (iv) a consistency check on Migdal’s approximation when strong e–e correlations coexist with sizable e-ph coupling. Together, these elements connect symmetry breaking, many-body screening, and higher-order scattering to the full hierarchy of transport coefficients and nonequilibrium response functions in twisted/stacked 2D materials.

Addressing these challenges will complete a rigorous microscopic framework for e-ph dynamics in low-dimensional Dirac systems, one that consistently links symmetry breaking, many-body screening, and higher-order scattering to the full hierarchy of transport coefficients and nonequilibrium response functions.

\section{Conclusion}


Electron-phonon scattering is the common microscopic thread that governs how heat is carried and ultimately dissipated, whether the host is a 3D metal, a band-gap semiconductor, or a 2D Dirac crystal. It sets electron lifetimes in metals, phonon lifetimes in semiconductors and 2D Dirac systems, and can even set the branch-dependent damping that decides which phonons survive in graphene. By combining density functional theory, density-functional perturbation theory, Wannier interpolation, and (when needed) \emph{coupled} electron-- and phonon--Boltzmann transport, and by treating screening beyond the Thomas--Fermi approximation, we now have a predictive, parameter free framework that captures e-ph coupling from mode-resolved phonon lifetimes in \(d\)-band metals to dynamically screened, higher-order channels in graphene nanoribbons.

Across the metallic regime, first-principles calculations revise the long-held view that lattice heat flow is negligible: once e-ph scattering is included, phonons can account for up to \(\sim\!40\%\) of \(\kappa_{\mathrm{tot}}\), with the magnitude set primarily by \(N(E_{\mathrm F})\), deformation potentials, and phonon frequencies. In semiconductors, solving the electron and phonon BTEs on equal footing resolves phonon drag, electron drag, and ultrafast non-equilibria, bringing theory into quantitative agreement with TDTR and isotope-controlled benchmarks (e.g., doped Si, GaAs, and polar nitrides). In 2D Dirac materials, symmetry, strain, carrier density, and finite size jointly establish a branch-selective hierarchy: in pristine suspended graphene, first-order ZA/ZO coupling is symmetry-forbidden, whereas substrates, edges, or tension activate/modify flexural channels and can shift the balance between in-plane and out-of-plane heat carriage. Moreover, when \(E_{\mathrm F}<<k_B T\), higher-order processes become essential in 2D Dirac crystals, exposing a rich arena for engineering thermal rectifiers through controlled symmetry breaking and higher-order coupling. This opens a rich arena for engineering thermal rectification via controlled symmetry breaking and tailored higher-order e-ph coupling. Table~2 synthesizes, across the material classes treated here, the dominant \emph{phonon} scattering mechanisms and the key control parameters governing lattice thermal conductivity at $T\simeq 300$–$500\,\mathrm{K}$.

\begin{table}[h]
\centering
\caption{A synthesized summary of dominant \emph{phonon} scattering channels and controlling levers at $T\!\approx\!300$–$500$\,K. Entries reflect trends compiled in this review (see Secs.~3–5).}
\begin{tabular}{p{3.6cm} p{6.1cm} p{5.8cm}}
\toprule
\textbf{Material class} & \textbf{Dominant channels at 300–500 K} & \textbf{Why (key levers / modeling notes)}\\
\midrule
\textbf{Metals} (noble $\to$ $d$-band; intermetallics)
&
e–ph sets leading \emph{phonon} lifetimes; 3ph anharmonicity captures intrinsic lattice resistance; $\kappa_{\mathrm{ph}}$ share: often $<\!10\%$ (nobles), can reach $\sim\!10$–$40\%$ in several $d$-band/intermetallic systems (once e–ph is included)
&
High $N(E_F)$ and screening shorten phonon MFPs ($\sim$ few–tens nm); optical branches in multi-atom bases add channels. \emph{Notes:} Boundary/size effects matter for thin films.\\
\midrule
\textbf{Semiconductors} (covalent vs polar)
&
3ph Umklapp dominates $\kappa_{\mathrm{ph}}$ (LA/TA in covalent; plus polar LO–TO coupling effects on phonon scattering in polar crystals); isotope and boundary scattering can be non-negligible. e–ph damping of \emph{phonons} becomes relevant at high carrier densities
&
Bonding character (short- vs long-range), carrier density, dielectric environment (screening). \emph{Notes:} At 300 K, coupled e–ph BTE mainly impacts Seebeck/drag rather than $\kappa_{\mathrm{ph}}$ in many cases.\\
\midrule
\textbf{2D Dirac} (graphene and kin)
&
With mirror symmetry $\sigma_h$ intact, selection rules suppress ZA couplings, leading to  LA/TA dominance; ZA is also suppressed in certain Dirac crystals such as graphene nanoribbons; substrates/gating/tension activate ZA channels and modify screening; nonadiabatic effects can arise at high $n$ or near flat-band limits
&
Symmetry selection rules (ZA protection), environment-dependent screening (gate/encapsulation), strain/tension (ZA linearization). \emph{Notes:} higher order in-plane scattering becomes important for 2D Dirac crystals at higher $T$; coherence (unified) term is usually negligible in simple monolayers.\\
\midrule
\textbf{Moiré/twisted graphene}
&
Enhanced e–ph in flat minibands; additional “moiré-Umklapp’’ channels; moiré phonons restructure phase space; coherences may contribute when inter-branch splittings $\sim$ linewidths
&
Flat-band DOS and modified screening; moiré-phonon mini-bands; remote interfacial phonons (substrates). \emph{Notes:} Start with Peierls–BTE; include coherence (unified) corrections when branches are quasi-degenerate; assess 4ph at high $T$.\\
\bottomrule
\end{tabular}
\end{table}

Looking forward, the path is clear: enlarge \emph{ab initio} datasets to include correlated and magnetic Dirac sheets; embed fully non-local, frequency-dependent screening in van der Waals heterostructures so that strong e–ph coupling is treated beyond the Thomas–Fermi limit; incorporate second-order (four-particle) scattering—including EP\,$\leftrightarrow$\,E$^{\ast}$P$^{\ast}$ channels—into coupled electron- and phonon-BTE solvers; and couple these strong-coupling formalisms to femtosecond pump–probe experiments for real-time validation. In the near term, practitioners should replace Thomas-Fermi with \emph{static/dynamic Lindhard} screening in low-density 2D Dirac systems, include \emph{four-phonon} processes in stiff/ultrahigh-$\kappa$ lattices and at elevated temperatures, deploy \emph{coupled} e–ph BTE where drag/Onsager consistency matters, and report XC choice, $k/q$-mesh densities, Wannierization quality, mean-free-path spectra, and boundary/substrate (remote-phonon) treatments to ensure reproducibility. Over the longer horizon, incorporating \emph{coherences/unified} corrections when branches are quasi-degenerate, treating nonadiabatic effects in flat-band moiré systems and validating against ultrafast probes, extending to correlated materials with DFT+$U$/DFT+DMFT and consistent screening, and building automated DFPT$\to$EPW workflows that integrate dynamic screening, four-phonon scattering, and (where needed) coupled e–ph BTE with uncertainty quantification and data release, will transform electron–phonon theory from a diagnostic tool into a genuinely predictive, strong-interaction design platform—capable of atomically precise control of heat and charge in next-generation electronic, photonic, and quantum devices.

\centerline{***}
\paragraph*{\bf Acknowledgments}
We thank the Perimeter Institute for Theoretical Physics for providing a stimulating research environment in which this work was carried out. We acknowledge the Anishinaabek, Haudenosaunee, Lūnaapéewak and Attawandaron peoples, on whose traditional lands Western University is located.

\bibliographystyle{unsrt}
\begin{footnotesize}
\bibliography{PINN}
\end{footnotesize}



\end{document}